\documentclass[nofootinbib,aps,prl,twocolumn,floats,floatfix,amsmath,amssymb,longbibliography,superscriptaddress,preprintnumbers]{revtex4-1}
\usepackage{graphicx}
\usepackage{dcolumn}
\usepackage{bm}
\usepackage{amsmath}
\usepackage{amsfonts} 
\usepackage{latexsym}
\usepackage{bbm}
\usepackage{color}
\usepackage{amssymb}
\usepackage{amsthm}
\usepackage{epsf}
\usepackage{epsfig}
\usepackage{caption3}
\usepackage{appendix}
\usepackage{cases}
\usepackage{subfigure}
\usepackage{multirow}
\usepackage[inkscapelatex=false]{svg}
\usepackage[colorlinks,linkcolor=blue,anchorcolor=blue,citecolor=blue,urlcolor=blue,]{hyperref}
\makeatletter

\newcommand{\Rmnum}[1]{\expandafter\@slowromancap\romannumeral #1@}
\makeatother

\begin{document}

\title{Self-consistent prediction of gravitational waves from cosmological phase transitions}

\author{Xiao Wang}%
\email{xiao.wang1@monash.edu}
\affiliation{School of Physics and Astronomy, Monash University, Melbourne, Victoria 3800, Australia}
\author{Chi Tian}%
\email{\textcolor{black}{Corresponding author:} ctian@ahu.edu.cn}
\affiliation{School of Physics and Optoelectronics Engineering, Anhui University, 111 Jiulong Road, Hefei, Anhui 230601, China}
\author{Csaba Bal\'azs}%
\email{csaba.balazs@monash.edu}
\affiliation{School of Physics and Astronomy, Monash University, Melbourne, Victoria 3800, Australia}



\begin{abstract}
Gravitational waves from cosmological phase transitions are novel probes of fundamental physics, making their precise calculation essential for revealing various mysteries of the early Universe.
In this work we propose a framework that enables the consistent calculation of such gravitational waves sourced by sound waves. 
Starting from the Lagrangian, this framework integrates the calculation of the dynamics of first-order phase transitions in a self-consistent manner, eliminating various approximations typically introduced by conventional methods.
At the heart of our approach is the congruous evaluation of the phase transition hydrodynamics that, at every step, is consistently informed by the Lagrangian. 
We demonstrate the application of our framework using the SM+$|H|^6$ model, deriving the corresponding gravitational wave spectrum. 
Our framework establishes a robust foundation for the precise prediction of gravitational waves from phase transitions.
\end{abstract}


\maketitle

\emph{\it \textbf{Introduction}}---
Nearly a decade after the first detection of gravitational waves (GWs) by LIGO and Virgo~\cite{LIGOScientific:2016aoc}, we are now on the brink of another major milestone: the detection of the stochastic gravitational wave background (SGWB). 
Preliminary evidence of this may have already been seen from recent Pulsar Timing Array experiments~\cite{NANOGrav:2023gor,Xu:2023wog,EPTA:2023fyk,Reardon:2023gzh}.
Phase transition gravitational waves (PTGWs), produced by cosmological first-order phase transitions (FOPTs), represent a significant source of SGWB and have been the focus of extensive studies in recent years. 
In particular, the possibility of an electroweak FOPT is of considerable interest due to its generation of an SGWB peaking around mHz, which is detectable by several proposed space-based experiments, such as LISA~\cite{LISA:2017pwj,Colpi:2024xhw}, TianQin~\cite{TianQin:2015yph}, Taiji~\cite{Hu:2017mde}, BBO~\cite{Corbin:2005ny}, DECIGO~\cite{Kawamura:2011zz}, and Ultimate-DECIGO~\cite{Kudoh:2005as}. 
Furthermore, FOPTs are associated with various extensions of the Standard Model (SM) of particle physics, making PTGWs a novel probe for new physics.

In order to pin down physics beyond the SM, precisely quantifying the associated PTGWs is essential but challenging. 
During a phase transition, GWs can be produced via three distinct mechanisms: bubble collisions, sound waves and  turbulence. 
Recent studies~\cite{Caprini:2015zlo,Caprini:2019egz,Athron:2023xlk} suggest that sound waves are usually the dominant source of PTGWs for thermal FOPTs, for which the scalar+fluid lattice simulations~\cite{Hindmarsh:2013xza,Hindmarsh:2015qta,Hindmarsh:2017gnf} are commonly employed to precisely derive the resulting GWs.
These simulations require a large simulation volume to accommodate hundreds of bubbles and fine grid spacing to resolve the bubble wall thickness, making them numerically expensive and time-consuming. 
Additionally, the wall velocity $v_w$ is adjusted by introducing a friction term, a free parameter, into the Higgs equation of motion (EoM).

Meanwhile, the majority of current studies utilise fitting formulae derived from lattice simulations, wherein all hydrodynamic quantities, including the kinetic energy fraction, are estimated by matching the effective potential of a specific particle physics model to a simplified equation of state (EoS), such as the bag model, through certain phase transition parameters, such as the bubble wall velocity $v_w$ and the strength parameter $\alpha$, etc.
In practice, the bubble wall velocity is left as a free parameter.
Depending on $v_w$, however, the broken power-law fitting formulae tend to deviate from various current analyses \cite{Jinno:2020eqg,Jinno:2022mie,Hindmarsh:2016lnk,Hindmarsh:2019phv,RoperPol:2023dzg,Sharma:2023mao,Giombi:2024kju}. 
These studies suggest that GW spectra generated by sound waves may more closely resemble to double broken-power law and even exhibit more complex features at low frequencies.

To address these challenges, alternative simplified numerical or semi-analytical methods have been proposed, including the hybrid simulation~\cite{Jinno:2020eqg,Jinno:2021ury}, the Higgsless simulation~\cite{Jinno:2022mie}, and the sound shell model~\cite{Hindmarsh:2016lnk,Hindmarsh:2019phv,Guo:2020grp,Wang:2021dwl,Cai:2023guc,RoperPol:2023dzg,Giombi:2024kju}. These approaches aim to enhance efficiency while maintaining accuracy. However, all of these methods rely on simplified models of EoS, such as the bag model, and a manually specified wall velocity.  These approximations introduce theoretical uncertainties into the GW spectrum.

\begin{figure*}[t]
	\centering
	\includegraphics[width=0.8\textwidth]{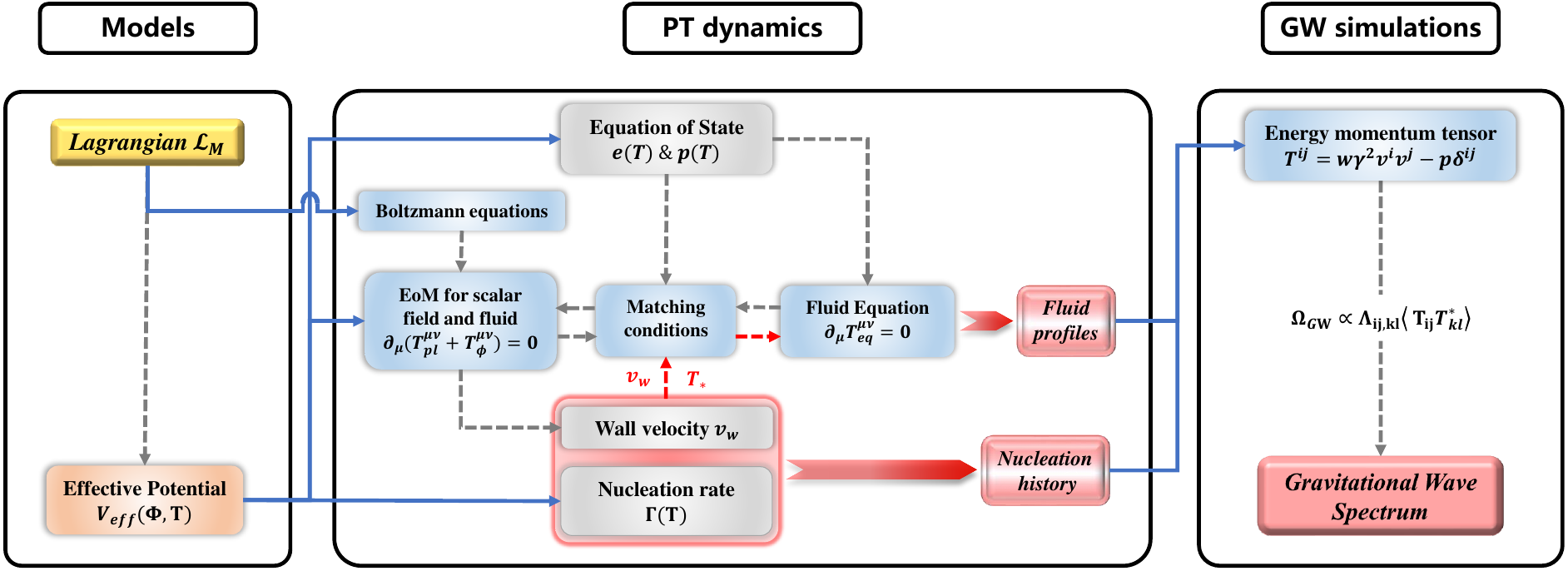}
	\caption{A schematic flowchart of our calculation framework for phase transition gravitational waves generated by sound waves.}
	\label{fig:Calf}
\end{figure*}

In this letter, to overcome these drawbacks of various methods, we propose a novel and efficient hydrodynamic-simulation-based framework capable, in principle, of quantifying PTGWs deterministically for any specific particle physics model.
As illustrated in the flowchart in Fig.~\ref{fig:Calf}, this method starts with the Lagrangian of a given model, from which the effective potential $V_{\rm eff}$ is derived. 
Utilising this effective potential, the EoS and nucleation rate $\Gamma$ are directly obtained, enabling the determination of the reference temperature $T_*$ from $\Gamma$. 
With the EoS and reference temperature $T_*$, the bubble wall velocity $v_w$ and the hydrodynamics prior to bubble collisions can be simultaneously determined by solving the EoM for scalar field and fluid.
By generalising the hybrid simulation, our method can then access the hydrodynamics after bubble collisions for more complicated and realistic EoS obtained by the effective potential. 
Finally, combining the nucleation history, determined by the nucleation rate and wall velocity, with hydrodynamics, the GW spectrum can be extracted.
We validate the applicability of our approach by conducting the first deterministic computation of GW spectra for the SM+$|H|^6$ model.

\emph{\it \textbf{Particle Physics Model}}---
We start from the SM+$|H|^6$ model, which introduces a dimension-6 operator $|H|^6$ to the Lagrangian of the SM~\cite{Zhang:1992fs,Grojean:2004xa,Chung:2012vg,Huang:2016odd,Wang:2020jrd}. 
The Higgs potential can then be extracted from the effective Lagrangian as 
\begin{equation}
    V(H) = \mu \left(H^\dagger H\right) + \lambda \left(H^\dagger H\right)^2 + \frac{1}{\Lambda^2}\left(H^\dagger H\right)^3\,,
    \label{eq:dim6}
\end{equation}
where the vacuum expectation value of the Higgs Doublet $H$ is $v_{\rm ew}\equiv \langle H \rangle  \approx 246.22~\mathrm{GeV}$, $\Lambda$ denotes the effective cutoff scale, $\lambda = \lambda_{\rm SM}\left(1 - \Lambda_{\rm max}^2/\Lambda^2\right)$ and $\mu^2 = \mu_{\rm SM}^2\left(-1 + \Lambda_{\rm max}^2/2\Lambda^2\right)$.
Here, $\Lambda_{\rm max}=\sqrt{3}v_{\rm ew}^2/m_h$ with the Higgs mass $m_h \approx 125~\mathrm{GeV}$, 
$\lambda_{\rm SM}$ and $\mu_{\rm SM}$ are the corresponding SM parameters.
According to the SM effective field theory, this effective model can represent general characteristics of many new physics models, such as the singlet extension~\cite{Espinosa:2007qk,Profumo:2007wc,Espinosa:2011ax,Chiang:2018gsn,Athron:2022jyi}, two-Higgs Doublet model~\cite{Lee:1973iz,Branco:2011iw,Dorsch:2016nrg,Basler:2017uxn,Fontes:2017zfn,Wang:2019pet}, etc. 
We initiate the study of FOPTs by deriving the effective potential $V_{\rm eff}(\phi, T)$ of this model.
In the unitary gauge, 
the tree level potential is written as
\begin{equation}
V_{\rm tree}(\phi) = \frac{1}{2}\mu^2\phi^2 + \frac{1}{4}\lambda\phi^4 + \frac{1}{8\Lambda^2}\phi^6\,.
\end{equation}
For simplicity, we approximate the finite-temperature effective potential by adding the thermal mass $cT^2\phi^2/2$ of $H$ to the tree level potential $V_{\rm tree}(\phi)$, thus
\begin{equation}
V_{\rm eff}(\phi, T) \approx -\frac{1}{3}aT^4 + \frac{\mu^2 + cT^2}{2}\phi^2 + \frac{\lambda}{4}\phi^4 + \frac{1}{8\Lambda^2}\phi^6\,,
\end{equation} 
where $a = g_*\pi^2/30$, $g_* \approx 106.75$ is the number of relativistic degrees of freedom at the electroweak phase transition, and $c = \frac{1}{16}\left(g'^2 + 3g^2 + 4y_t^2 + 4\frac{m_h^2}{v_{\rm ew}^2} - 12\frac{v_{\rm ew}^2}{\Lambda^2}\right)$.  Here $g'$ and $g$ are the $U(1)_Y$ and $SU(2)_L$ gauge couplings, and $y_t$ is the top quark Yukawa coupling.
Note that we keep the pure temperature-dependent term $-aT^4/3$ in the effective potential, as it is necessary for the hydrodynamics of the phase transition.

\emph{\it \textbf{Phase Transition Dynamics}}---
The dynamics of a thermal first-order phase transition can be, in practice, divided into two parts: the dynamics of the bubble wall, and the hydrodynamics of the plasma.
To be able to analyse the dynamics of these two parts, we need to resolve both the EoS and the bubble nucleation rate from the given model.
In the thermodynamic limit, the free energy is $\mathcal{F}(T) = V_{\rm eff}(\phi_m(T),T)$~\cite{Laine:2016hma}.
Hence, with the effective potential, we can derive the EoS of the SM+$|H|^6$ as
\begin{equation}
\begin{split}
    &p_+(T) = \frac{1}{3}aT^4\,,\quad e_+(T) = aT^4\,,\\
     &p_-(T) = \frac{1}{3}aT^4 - \frac{\mu^2 + cT^2}{2}\phi_{m}^2 - \frac{\lambda}{4}\phi_{m}^4 - \frac{\kappa}{8\Lambda^2}\phi_{m}^6\,,\\
    &e_-(T) = aT^4 + \frac{\mu^2 - cT^2}{2}\phi_{m}^2 + \frac{\lambda}{4}\phi_{m}^4 + \frac{\kappa}{8\Lambda^2}\phi_{m}^6\,,
\end{split}
\end{equation}
where $p_\pm$ and $e_\pm$ represent pressure and energy density respectively, subscripts $\pm$ denotes the high-temperature phase and the low-temperature phase, and $\phi_m(T)$ is the global minimum of the effective potential at temperature $T$.
The enthalpy of the system is $w_\pm = e_\pm + p_\pm$.

With the effective potential, one can also derive the nucleation rate $\Gamma(t) = \Gamma_0 \exp[S_E(t)]$, where $S_E$ denotes the bounce action which could be calculated by solving the bounce equation~\cite{Wang:2020jrd}.
In this work, to simplify the calculation, we parameterise the bubble nucleation rate as~\cite{Enqvist:1991xw,Jinno:2021ury}
\begin{equation}
    \Gamma(t) \approx \beta_*^4 \exp{[\beta_*(t - t_*)]}\,,
    \label{eq:nucl}
\end{equation}
where $\beta_* = \frac{dS_E}{dt}\big|_{t=t_*}$, and the reference time $t_*$ is the time corresponds to the reference temperature $T_*$.
Here, for simplicity, the reference temperature is chosen as the nucleation temperature $T_n$ obtained by the conventional method~\cite{Wang:2020jrd}.
In addition to the nucleation rate, understanding the bubble wall velocity~\cite{Moore:1995ua,Moore:1995si,Laurent:2020gpg,Laurent:2022jrs,Wang:2020zlf,Jiang:2022btc}, which characterises the wall dynamics, is crucial for constructing the full nucleation history.
In our framework, the bubble wall velocity is assumed to be a constant and to maintain its steady-state value.
Understanding the full nucleation history is also important for hydrodynamics.

Next, we look for ways to determine the steady-state bubble wall velocity $v_w$. Since $v_w$ is strongly related to the single-bubble hydrodynamics, it must be determined simultaneously while determining the hydrodynamics of a single bubble. 
The computation begins with the energy-momentum tensor (EMT) of the phase transition system.
In general, the EMT consist of the scalar field and the plasma:
\begin{equation}
    T^{\mu\nu} = T_\phi^{\mu\nu} + T_{\rm pl}^{\mu\nu}\,. 
\end{equation}
Here the EMT of the scalar field is 
\begin{equation}
T_\phi^{\mu\nu}  = \partial^\mu\phi \partial^\nu\phi - g^{\mu\nu}\left[\frac{1}{2}\partial_\alpha\phi\partial^\alpha\phi - V_0(\phi)\right]\,,
\end{equation}
with $V_0(\phi)$ being the zero-temperature part of the effective potential, and $T_{\rm pl}^{\mu\nu}$ representing the EMT of the plasma
\begin{equation}
    T_{\rm pl}^{\mu\nu} = \sum_i\int\frac{d^3\tilde{\mathbf{p}}}{(2\pi)^3E_{\tilde{p}}}\tilde{p}^\mu \tilde{p}^\nu f_i \,.
\end{equation}
Above, $f_i$ is the distribution function of the particle species $i$, $\tilde{p}^\mu\equiv(E_{\tilde{p}},\tilde{\mathbf{p}})$ is the 4-momentum.
In both phases, the plasma is in thermal equilibrium and dominates the EMT, thus the scalar field part is negligible.
However, across the bubble wall, the contribution of the scalar field should be taken into account, and the plasma becomes out of equilibrium.
Assuming the deviation from equilibrium is small, $f_i \approx f_i^{\rm eq} + \delta f_i$, with $f_i^{\rm eq} = 1/(\exp[\tilde{p}_\mu u^\mu/T] \pm 1)$, where $u^\mu$ is the 4-velocity of the plasma. 
The EMT of the plasma is then the sum of the equilibrium and the out-of-equilibrium parts, i.e., $T_{\rm pl}^{\mu\nu} = T_{\rm eq}^{\mu\nu} + T_{\rm oeq}^{\mu\nu}$.
Here we assume that the plasma is a perfect fluid: $T_{\rm eq}^{\mu\nu} = (e + p) u^\mu u^\nu - pg^{\mu\nu}$.
Based on Lorentz invariance and the symmetry, the out-of-equilibrium EMT can be constructed as $T_{\rm oeq}^{\mu\nu} = T_{\rm oeq, g}g^{\mu\nu} + T_{\mathrm{oeq}, u}$~\cite{Laurent:2022jrs}, with
\begin{equation}
\begin{split}
T_{\rm oeq, g} =& \frac{1}{2}\sum_i(m_i^2\Delta_{00}^i + \Delta_{02}^i - \Delta_{20}^i)\,,\\
T_{\mathrm{oeq}, u} =& \frac{1}{2}\sum_i\Big[(3\Delta_{20}^i - \Delta_{02}^i - m_i^2\Delta_{00}^i)u^\mu u^\nu \\
&+ (3\Delta_{02}^i - \Delta_{20}^i + m_i^2\Delta_{00}^i)\bar{u}^\mu\bar{u}^\nu\\
&+ 2\Delta_{11}^i(u^\mu\bar{u}^\nu + \bar{u}^\mu u^\nu) \Big]\, .
\end{split}
\label{eq:EoS}
\end{equation}
The 4-velocity satisfies $\bar{u}^\mu u_\mu = 0$, and
\begin{equation}
\Delta_{mn}^i \equiv \int\frac{d^3\tilde{\mathbf{p}}}{(2\pi)^3E_{\tilde{p}}}(\tilde{p}_\mu u^\mu)^m(-\tilde{p}_\mu\bar{u}^\mu)^n\delta f_i\,.
\end{equation}
Therefore, in the wall frame, with planar approximation, the total energy-momentum conservation $\partial_\mu T^{\mu\nu}$ at steady state implies
\begin{subequations}
\begin{align}
    \partial_z^2\phi + \frac{\partial V_{\rm eff}}{\partial \phi} + \sum_i\frac{\partial(m_i^2)}{\partial\phi} \frac{\Delta_{00}^i}{2} &= 0\,,\label{eq:EOM}\\
    \partial_z\left[w\gamma^2v + T_{\rm oeq}^{30}\right] &= 0\,,\label{eq:EF1}\\
    \partial_z\left[\frac{1}{2}(\partial_z\phi)^2 - V_{\rm eff} + w\gamma^2v^2 + T_{\rm oeq}^{33}\right] &= 0 \,.\label{eq:EF2}
\end{align}
\label{eq:BFEOM}
\end{subequations}
Considering the out-of-equilibrium nature of the plasma EMT, eqs.~\eqref{eq:BFEOM} need to be solved simultaneously with the Boltzmann equation to obtain $\delta f_i$ for every relevant particle species $i$ along with $\Delta_{mn}^i$, which is
\begin{equation}
    \left[\tilde{p}_z\partial_z - \frac{1}{2}\partial_z(m_i^2)\partial_{\tilde{p}_z}\right]\delta f_i = \mathcal{C}^{\rm lin} + \mathcal{S}_i\,,
    \label{eq:Boltz}
\end{equation}
where the source term $\mathcal{S}_i$ is given by the action of the operator $\hat{\mathcal{D}}\equiv\left[\tilde{p}_z\partial_z - \partial_z(m_i^2)\partial_{\tilde{p}_z}/2\right]$ on the equilibrium distribution function $f_i^{\rm eq}$. 
The linearised collision terms $\mathcal{C}^{\rm lin}$ can be computed by
\begin{subequations}
\begin{align}
    \mathcal{C}^{\rm lin}[f]&=\sum_a\frac{1}{2N_{\tilde{p}}}\int\frac{d^3\tilde{\mathbf{k}}d^3\tilde{\mathbf{p}}'d^3\tilde{\mathbf{k}}'}{(2\pi)^92E_{\tilde{k}}2E_{\tilde{p}'}2E_{\tilde{k}'}}  |\mathcal{M}_a|^2  (2\pi)^4\notag\\
&\times\delta^4(\tilde{p}+\tilde{k}-\tilde{p}'-\tilde{k}')\mathcal{P}^{\rm lin}[f]\,,\\
\mathcal{P}^{\rm lin}[f] &=  f_i^{\rm eq}(\tilde{p})f_j^{\rm eq}(\tilde{k})f_{l}^{\rm eq}(\tilde{p}')f_{m}^{\rm eq}(\tilde{k}')\bigg[\frac{e^{\mathcal{E}_{\tilde{k}}}\delta f_i(\tilde{p})}{f_i^2(\tilde{p})} + \notag\\
&\frac{e^{\mathcal{E}_{\tilde{p}}}\delta f_j(\tilde{k})}{f_j^2(\tilde{k})} - \frac{e^{\mathcal{E}_{\tilde{k}'}}\delta f_l(\tilde{p}')}{f_l^2(\tilde{p}')}
- \frac{e^{\mathcal{E}_{\tilde{p}'}}\delta f_m(\tilde{k}')}{f_m^2(\tilde{k}')} \bigg]\, .
\end{align} 
\end{subequations}
Here $\tilde{p}$ denotes the momentum of the incoming particle for which $\delta f$ is computed, with $N_{\tilde{p}}$ representing its degrees of freedom. The momentum of another incoming particle is $\tilde{k}$, while $\tilde{p}'$ and $\tilde{k}'$ are the momenta of the outgoing particles.
The summation is over all relevant scattering amplitudes $|\mathcal{M}_a|^2$ derived at leading-log order \cite{Moore:1995si,Laurent:2020gpg,Laurent:2022jrs,Jiang:2022btc}. 
In this work, we only consider the contribution of the top quark to the the friction term in eq.\eqref{eq:EOM}, so the relevant scattering processes include $\bar{t}t\to gg$, $tg\to tg$, and $tq\to tq$.  Here $g$ denotes gluons and $q$ represents other quarks, both of which are assumed to be in thermal equilibrium.
Additionally, indices $i$, $j$, $l$, and $m$ correspond to different particle species, and $\mathcal{E}_{\tilde{p}}\equiv \tilde{p}_\mu u^\mu$.
For simplicity $\mathcal{C}^{\rm lin}$ is calculated in the plasma frame.
The Chebyshev spectral method~\cite{boyd2013chebyshev,Laurent:2022jrs} is then employed to solve the Boltzmann equation. 

To determine $v_w$, boundary conditions of eqs.~\eqref{eq:BFEOM} must be applied.
These boundary conditions are related to the single-bubble hydrodynamics governed by the EoS~\eqref{eq:EoS}.
Integrating eqs.~\eqref{eq:BFEOM} across the wall gives the matching conditions~\cite{Espinosa:2010hh,Wang:2020nzm}
\begin{equation}
\begin{split}
    w_+\gamma_+^2v_+ &= w_-\gamma_-^2v_-\,,\\
   w_+\gamma_+^2v_+^2 + p_+ &= w_-\gamma_-^2v_-^2 + p_-\,.
\end{split}
\label{eq:match}
\end{equation}
Our subsequent hydrodynamic analysis shows that the SM+$|H|^6$ model allows for detonations ($v_w = |v_+|$), deflagrations ($v_w = |v_-|$), and hybrid modes ($v_- = c_{s,-}$, where $c_{s,-}$ is the sound speed just behind the wall).
More details on hydrodynamic analysis can be found in our companion paper~\cite{Long:2024}.
While the boundary conditions for the detonation mode can be directly derived from eq.~\eqref{eq:match}, those for the deflagration and hybrid modes require solving the fluid profiles between the bubble wall and the shock front.
Since the fluid profiles are away from the bubble wall, 
it can then be derived from the energy momentum conservation of the plasma in equilibrium, i.e., $\partial_\mu T^{\mu\nu}_{\rm eq} = 0$~\cite{Espinosa:2010hh,Wang:2020nzm}.
At steady state, for spherical bubbles, we thus have
\begin{equation}
\begin{split}
2\frac{v}{\xi} &= \gamma^2(1 - v\xi)\left[\frac{\mu^2}{c_s^2(T)} - 1\right]\partial_\xi v\,,\\
\partial_\xi T &= T\gamma^2\mu \partial_\xi v\, .
\end{split}
\label{eq:fluideqs}
\end{equation}
Here, $\mu(\xi, v) \equiv (\xi - v)/(1 - \xi v)$, the sound speed $c_s^2(T) = (dp/dT)/(de/dT)$, $\xi \equiv r/t$, where $r$ denotes the distance from the bubble centre and $t$ is the elapsed times since nucleation.
At the shock front, the conservation of the EMT also gives eqs.~\eqref{eq:match}, which determine the position of the shock front.
By applying the shooting method, we can thus find corresponding boundary conditions for deflagrations and hybrid modes.
We then employ the profile ansatz $\phi(z) = 0.5\phi_0[1 -\tanh(z/L_w)]$,  and two moments~\cite{Moore:1995si}:
$P = -\int dz [\mathrm{Eq}.\eqref{eq:EOM}] d\phi/dz$, and $G = \int dz [\mathrm{Eq}.\eqref{eq:EOM}] (2\phi/\phi_0 - 1)d\phi/dz$ to simplify the calculation of eqs.~\eqref{eq:BFEOM}, where $\phi_0=\phi_m(T_-)$ and $L_w$ is the bubble wall width.
With an iterative method~\cite{Laurent:2022jrs}, the steady-state bubble wall velocity $v_w$ can be eventually determined and the fluid profiles around a single bubble can also be derived simultaneously from eqs.~\eqref{eq:fluideqs}.
We show $v_w$, $L_w$ and other phase transition parameters for two cutoff scales in Table.~\ref{tb:PTp}. 
Depending on the cutoff scales, we find that our scheme predicts the existence of a deflagration and a detonation mode for benchmark model $\mathrm{BP}_1$ and $\mathrm{BP}_2$, respectively. 

\begin{table}[!t]
	\centering
	\footnotesize
	\setlength{\tabcolsep}{2.5pt}
	\renewcommand{\arraystretch}{1.2}
	\begin{tabular}{c|c|c|c|c|c} 
        \hline 
		\hline
		&$\Lambda\rm~[GeV]$  & $T_n~\mathrm{[GeV]}$ & $\beta_n/H_n$& $v_w$  & $L_wT_n$   \\ \hline 
        
		$\mathrm{BP}_1$ &740 & 95.58 & 17217 & 0.43 & 10.40 \\ 
        \hline
        
        $\mathrm{BP}_2$ &640 & 73.50 & 1806 &0.99995 &4.32\\ 
        \hline
        \hline
        
    \end{tabular}
    \caption{The phase transition parameters for two different benchmark points of the SM+$|H|^6$.}
    \label{tb:PTp}
\end{table}

\begin{figure}[!t]
    \centering
    \includegraphics[width=0.3\textwidth]{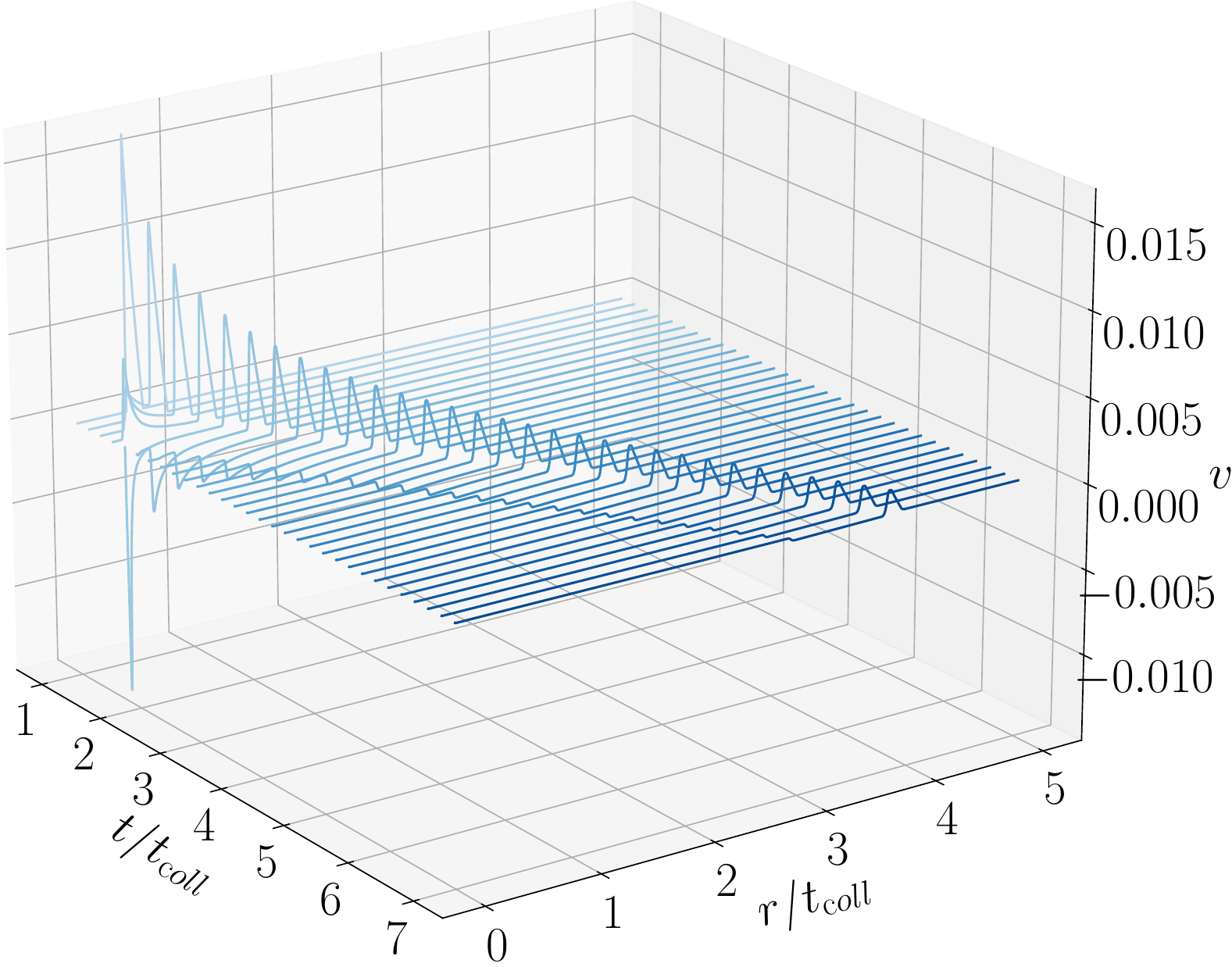}
    \includegraphics[width=0.3\textwidth]{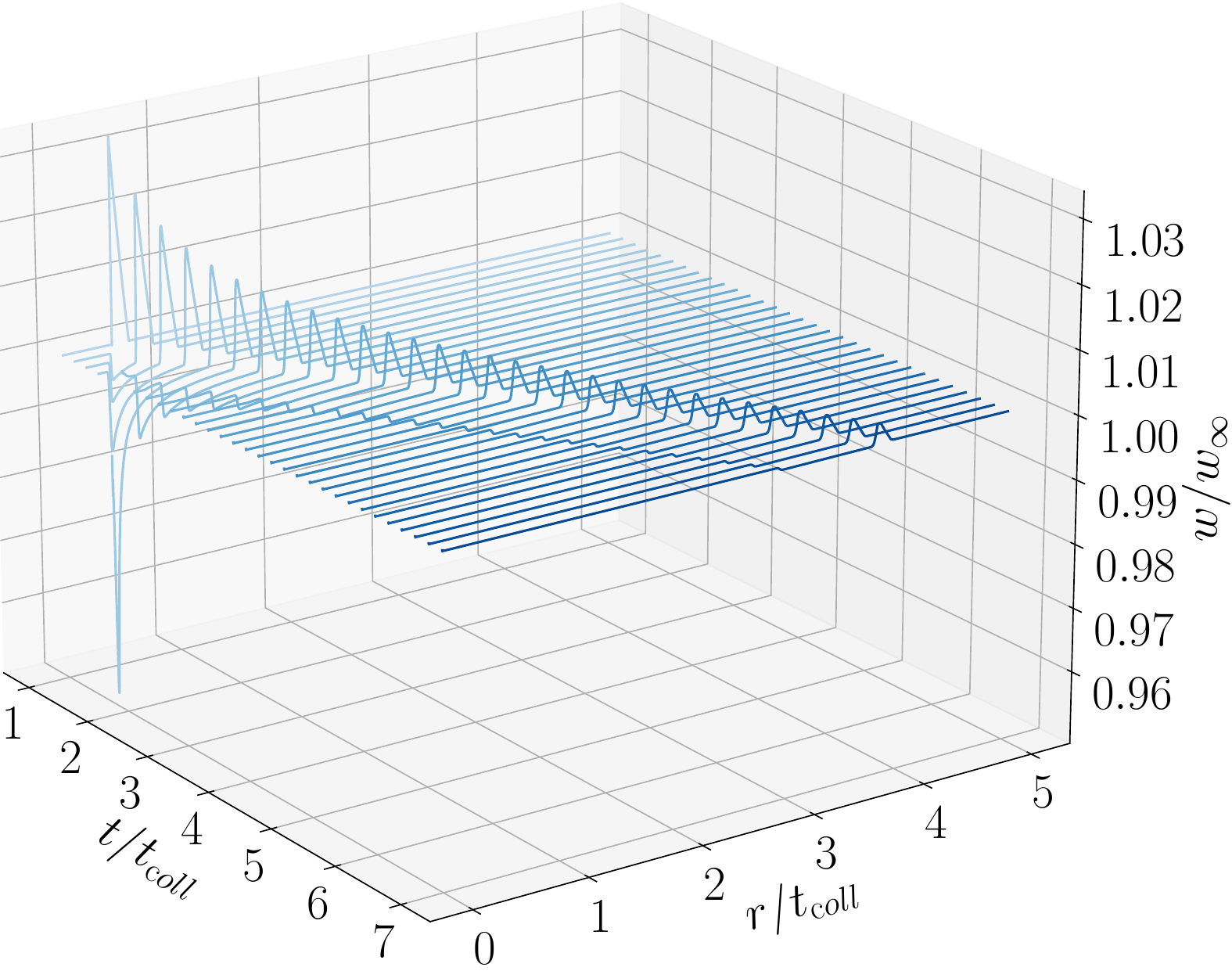}
    \caption{The evolution of spherically symmetric fluid profiles after collision for benchmark model $\mathrm{BP}_1$. \textbf{Top}: the fluid velocity profile. \textbf{Bottom}: the enthalpy profile. Here, $w_\infty = w_+(T_n)$ denotes the enthalpy deep inside the high-temperature phase.}
    \label{fig:fluid}
\end{figure}

Until now we have derived the important parameters to describe the dynamics before the bubble collision.
However, to compute gravitational waves, the hydrodynamics after the collision of bubble walls is essential.
To simplify the calculation, we assume the spherical symmetry is still preserved for each radial fluid profile of its corresponding wall surface element, and that the scalar field disappears soon after collisions.
The effect of the bubble collision is thus just simply shifting the temperature ahead of the fluid shell to the temperature deep inside the bubbles.
Based on energy-momentum conservation $\partial_\mu T^{\mu\nu}_{\rm eq} = 0$ and the spherical approximation, the equations describing the fluid evolution after collisions are
\begin{equation}
    \begin{split}
        \partial_t E + \partial_r Z &= -\frac{2}{r} Z\,,\\
        \partial_t Z + \partial_r [Zv + p] &= -\frac{2}{r}Zv\, ,
    \end{split}
    \label{eq:EZ}
\end{equation}
where $Z\equiv w\gamma^2v$ and $E\equiv w\gamma^2 - p$.
The Kurganov-Tadmor~\cite{Kurganov:2000ovy} scheme is used to evolve these equations.
We use $E + p - \left[w + \sqrt{w^2 + 4Z^2}\right]/2 = 0
$ to derive the temperature $T$ with given $E$ and $Z$.
Then with the determined $T$ and the EoS~\eqref{eq:EoS}, we derive $v$ from the definition of $Z$ with a root finding algorithm.
The evolution of fluid profiles of $\mathrm{BP}_1$ are shown in Fig.~\ref{fig:fluid}.

So far, we have explored the bubble wall dynamics and hydrodynamics both before and after bubble collisions.
Nevertheless, our hydrodynamical analysis only focused on the fluid shell around a single bubble before collision or a single radial fluid element after collision.
To access the full hydrodynamics of FOPTs, the hydrodynamics obtained so far should be combined with the full nucleation history.
We further assume the system to behave perturbatively, the hydrodynamic quantities could then be given by the superposition of different bubbles: 
\begin{equation}
\begin{split}
    \frac{\Delta w}{ w_0} (t, \vec{x}) &\simeq \sum_{i:\rm bubbles} \frac{\Delta w _{\vec{n}_i}}{w_0}(t, |\vec{x} - \vec{n}_i|)\,,\\
    \vec{v}(t, \vec{x}) &\simeq  \sum_{i:\rm bubbles}  \vec{v}_{\vec{n}_i}(t, |\vec{x} - \vec{n}_i|)\, .
\end{split}
\label{eq:Fh} 
\end{equation}
Here $w_0$ is the enthalpy deep inside the bubble, $\Delta  w _{\vec{n}_i} =  w _{\vec{n}_i}-w_0$, and $\vec{n}_i$ denotes the position vectors of nucleation points.
Then, as detailed in Ref.~\cite{Jinno:2020eqg,Long:2024}, once the collision time of every surface element $t_{\rm coll}^i$ is determined, the 1D spherical solutions can be projected onto a 3D Cartesian lattice as follows.  When $t<t_{\rm coll}$, the radial profile is self-similar and given by eqs.~\eqref{eq:fluideqs}, whereas when $t>t_{\rm coll}$, the radial profile is obtained from eqs.~\eqref{eq:EZ}. Ultimately, GWs can then be calculated from these 3D quantities given by~\eqref{eq:Fh}, constructed by 1D profiles.

\begin{figure}[t!]
    \centering
    \includegraphics[width=0.16\textwidth]{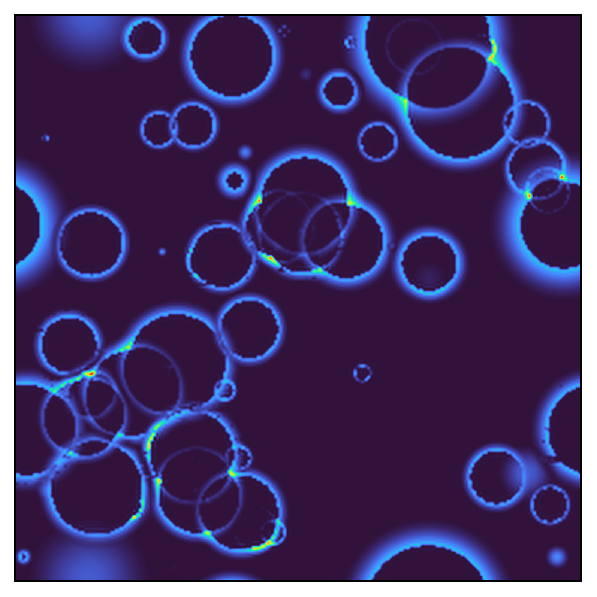}%
    \includegraphics[width=0.16\textwidth]{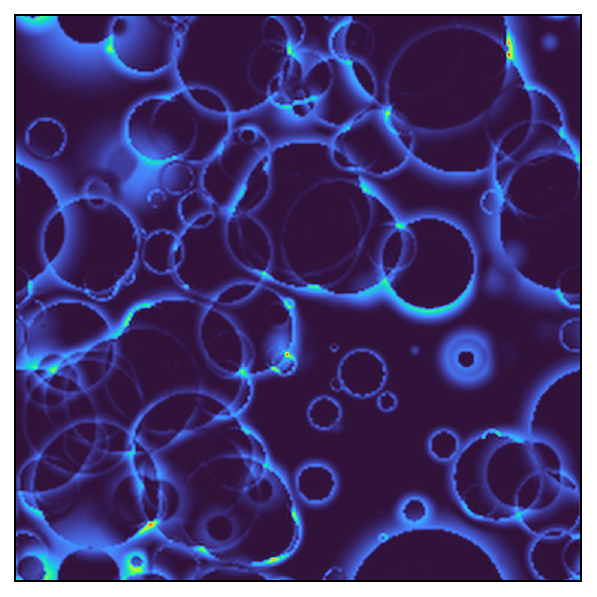}%
    \includegraphics[width=0.16\textwidth]{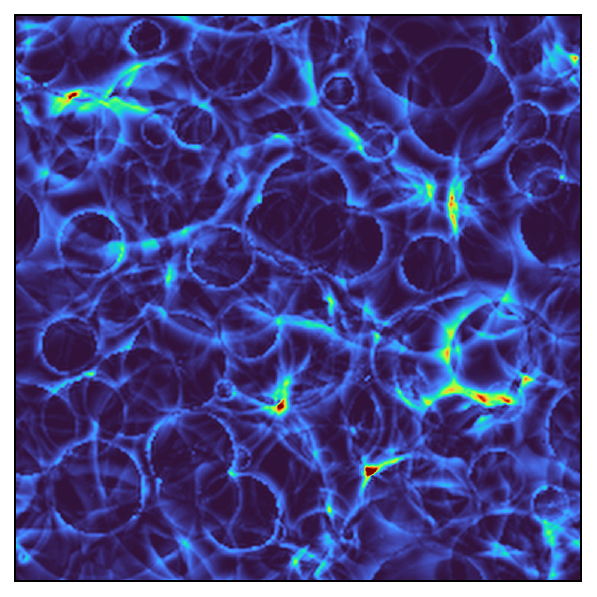}
    \caption{Snapshots of the time evolution of kinetic energy density $w \gamma^2 v^2 (t, \vec{x})$ in our simulations.  These figures correspond to a simulation for the benchmark parameter set $\mathrm{BP}_1$ in Table \ref{tb:PTp}, with a simulation volume $\mathcal{V} = L^3=(40 v_w / \beta_*)^3$ and grid resolution $N^3=256^3$. }
    \label{fig:bb}
\end{figure}

\emph{\it \textbf{GW Spectrum}}---
Gravitational waves are characterised by tensor perturbations $h_{ij}$ of the Friedmann-Robertson-Walker metric $ds^2 = -dt^2 + a^2(t)(\delta_{ij} + h_{ij})dx^idx^j$, sourced by $T^{\mu\nu}$ according to the linearised Einstein equation $\Box h_{ij} = 16\pi G \Lambda_{ij,kl}T^{kl}$, where $\Box$ and $G$ are the d'Alembertian and the Newtonian constant.
Given the short duration of the GW sources, the scale factor $a(t)$ can be neglected.
Hence, the GW spectrum $\Omega_{\rm GW}^*$ at production time can be derived by
\begin{equation}
\begin{split}
    \Omega_{\rm GW}^*(q) &\equiv  \frac{1}{\rho_{\rm crit}}\frac{d\rho_{\rm GW}}{d\ln k} 
    \approx \frac{4H^2\tau_{\rm sw}}{3\pi^2\beta}\frac{q^3\beta}{w_\infty^2\mathcal{V}\mathcal{T}}\\
    &\times\int\frac{d\Omega_k}{4\pi}\left[\Lambda_{ij,kl}T_{ij}(q,\vec{k})T^*_{kl}(q,\vec{k})\right]_{q=|\vec{k}|}\, .
\end{split}
\label{eq:GWpt}
\end{equation}
In this equation $q$ is the angular frequency, $\vec{k}$ the momentum of the GW waves, $\Lambda_{ij,kl}$ is the projection on the transverse-traceless part of the energy-momentum tensor $T_{ij}$, $\mathcal{V}$ the volume of the simulation box, and $\mathcal{T}$ the simulation time.
We use $\tau_{\rm sw}\approx R_*/\sqrt{K_{\rm fl}}$ to estimate the lifetime of the sound waves,
where $K_{\rm fl}$ is the kinetic energy fraction, which can be derived by the method given in Ref.~\cite{Wang:2023jto}, and the mean bubble separation $R_* = (8\pi)^{1/3}v_w/\beta_*$.

For the sound waves, the transverse-traceless part of the EMT is $T^{ij}(t, \vec{x}) = w(t, \vec{x})\gamma^2(t, \vec{x})v^i(t, \vec{x})v^j(t, \vec{x})$, 
which can be constructed from the projected enthalpy and velocity fields at different time described in eqs.~\eqref{eq:Fh}.
We demonstrate the 2D slices for the evolution of the kinetic energy density with the previously mentioned projection in Fig.~\ref{fig:bb}.
The GW spectrum $\Omega_{\rm GW}^*$ at the production time can then be obtained by plugging the momentum-space $T^{ij}$ into eq.~\eqref{eq:GWpt}.
To obtain the GW spectrum of today, we need to consider the cosmic expansion which yields
\begin{equation}
    \Omega_{\rm GW}(f) = 3.57 \times 10^{-5}\left(\frac{100}{g_*}\right)^{1/3}\Omega_{\rm GW}^*(q)\, .
\end{equation}
Above the redshifted frequencies are $f(q) = 2.6\times10^{-6}\mathrm{Hz}\frac{q}{\beta}\frac{\beta}{H_*}\frac{T}{100\mathrm{GeV}}\left(\frac{g_*}{100}\right)^{1/6}$,
and $H_*$ is the Hubble parameter obtained by the Friedmann equation at $T_*$.

\begin{figure}[!t]
    \centering
    \includegraphics[width=0.45\textwidth]{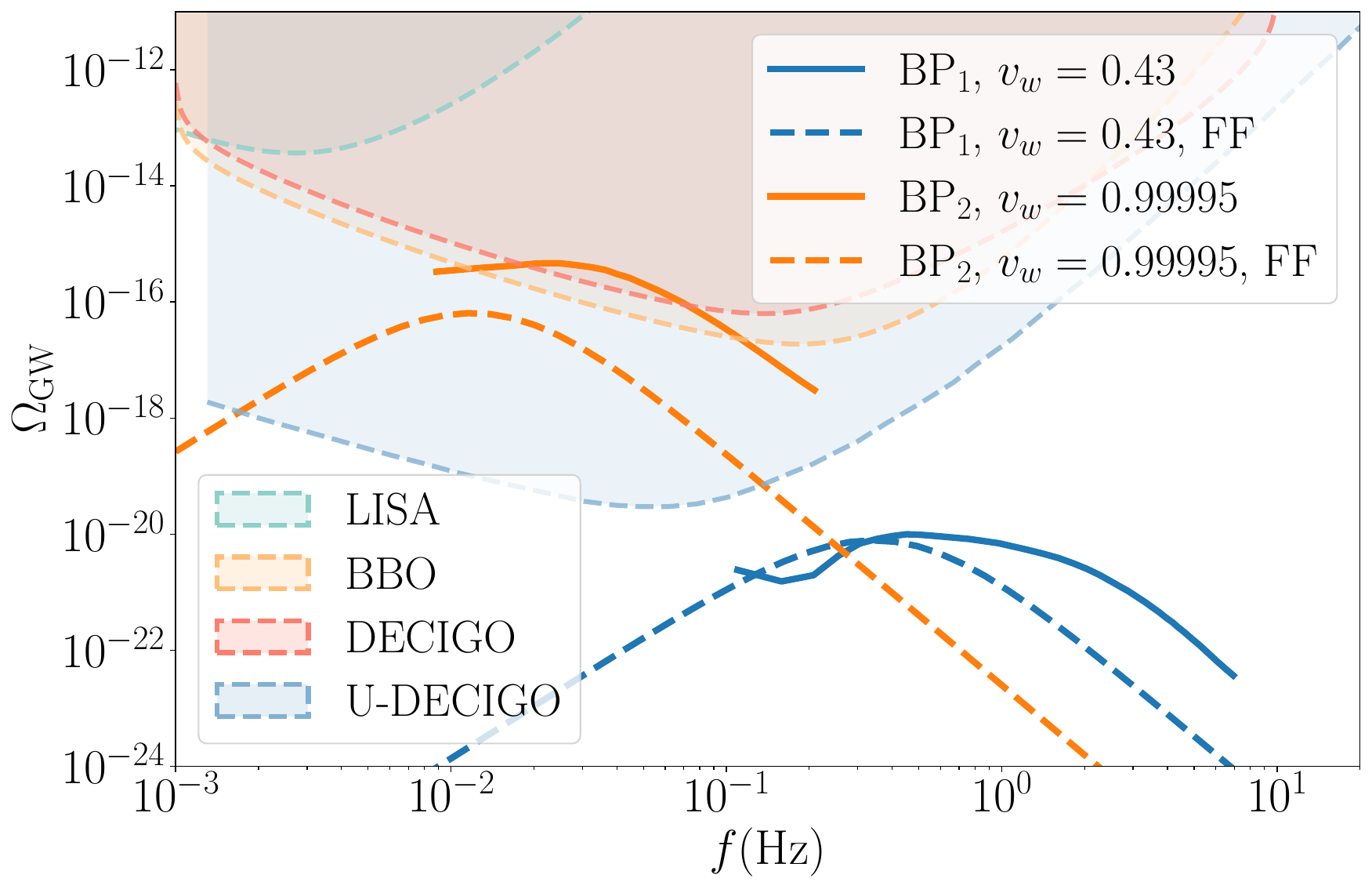}
    \caption{GW spectra for the benchmark models in Table \ref{tb:PTp}. The solid blue line and the solid orange line denote the GW spectra for $\mathrm{BP}_1$ and $\mathrm{BP_2}$, respectively, computed using our scheme. The dashed curves of the same colour  represent  spectra calculated using fitting formulae. The shaded regions represent the power-law-integrated sensitivity~\cite{Thrane:2013oya,Schmitz:2020syl} of various ongoing and planned GW detectors. }
    \label{fig:GWs}
\end{figure}
In Fig.~\ref{fig:GWs}, we show PTGW spectra  for the corresponding benchmark models listed in Table~\ref{tb:PTp}. 
We find that our scheme predicts that the spectrum of $\mathrm{BP}_1$ falls below the sensitivity of various future GW detectors. 
In contrast, the spectrum of $\mathrm{BP}_2$ is within the sensitivity range of BBO, DECIGO and Ultimate-DECIGO but remains outside of the sensitivity region of LISA. 
This can be attributed to the higher phase transition strength.  
Additionally, we present the result obtained from the fitting formulae of lattice simulations, depicted as dashed curves in Fig.~\ref{fig:GWs}. 
By comparing these spectra, we find that while the peak amplitude of the GW spectrum obtained using our method aligns with the fitting formulae for $\mathrm{BP}_2$, there is a significant discrepancy in amplitude across the accessible frequency range, with the spectral shapes derived from different methods also showing deviations. 
Additionally, our method suggests the presence of more complex features in the low-frequency range. 
For $\mathrm{BP}_2$, our approach yields a spectrum with an amplitude approximately an order of magnitude larger and a different shape compared to those predicted by the fitting formulae.
Therefore, these comparisons indicate that the conventional fitting formulae may not accurately capture the shape and amplitude of the GW spectrum.
Due to the limitations of our current computational resources, we are unable to compute the spectrum in the low- and high-frequency limit using our method, as these regimes require a larger simulation box and finer grid resolution.
Because of this, we only show our GW spectra in a limited frequency range.  
Outside of this range our present numerical calculation leads to an unphysical rise of the spectrum which indicates the range of its validity.

We also emphasise that although model $\mathrm{BP}_2$ possesses an ultra-relativistic bubble wall velocity, the sound wave remains the dominant source of GWs.  This can be demonstrated through the following brief analysis of the energy budget.
In general, the energy stored in the wall is $E_{\rm wall}(t) \approx 4\pi R_b(t)^2\gamma_w\sigma - \frac{4}{3}\pi R_b(t)^3 \Delta p$~\cite{Ellis:2019oqb,Ellis:2020nnr}, where $\sigma \approx \int dr [\frac{1}{2}(d\phi/dr)^2 + V_{\rm eff}]$ is the surface tension~\cite{Linde:1981zj,Laine:2016hma}, $R_b$ is the bubble radius, and $\Delta p$ is the net pressure acting on the wall.
Assuming the duration of the initial accelerated expansion is short, $R_b(t) = v_w t$.
Further, the net pressure should be zero at steady state, i.e. $\Delta p = 0$.
The energy fraction of the wall is defined as $K_{\rm wall} = E_{\rm wall}/E_{\rm tot}$, where $E_{\rm tot} = \frac{4\pi}{3}R_b(t_*)^3 e_+$.
Therefore, at the nucleation temperature, the energy fraction of the wall $K_{\rm wall}\sim10^{-10}$.
In addition, the kinetic energy fraction of the fluid shell is $K_{\rm fl} \sim 10^{-4}$. 
The sound wave is thus still the dominant source of PTGWs.

\emph{\it \textbf{Conclusion and Outlook}}--- Precise calculations of PTGWs are crucial for accessing new physics beyond the SM with future space-based GW experiments. 
However, besides theoretical uncertainties induced by the perturbative calculation for the effective potential, the conventional calculations of PTGWs suffer from other inconsistencies such as relying on fitting formulae derived under incompatible assumptions, employing a simplified treatment of the EoS, and leaving the bubble wall velocity as a free parameter.
To overcome these inconsistencies, we propose a hydrodynamic-simulation-based framework that allow us to perform the first consistent calculation of PTGWs sourced by the sound waves for the SM+$|H|^6$.
During the calculation of the phase transition hydrodynamics, our framework consistently relies on the Lagrangian instead of using approximations.
Compared with the results derived by the conventional method, our results show significant difference both in the spectral shape and amplitude.
In summary, our framework represents a significant advancement in the precise calculation of PTGWs. 
Future improvements could involve incorporating the latest refined calculations of the effective potential~\cite{Croon:2020cgk,Schicho:2022wty,Gould:2023ovu,Ekstedt:2024etx} based on dimensional reduction effective theory~\cite{Farakos:1994kx,Kajantie:1995dw}, which would enable more robust predictions of PTGWs.

\begin{acknowledgments}
\emph{\it \textbf{Acknowledgments}}--- X.W. would like to thank Benoit Laurent, Henrique Rubira, Michael Bardsley, and Lachlan Morris for enlightening and useful discussions.
C.T. is supported by the National Natural Science Foundation of China (Grants No. 12405048) and the Natural Science Foundation of Anhui Province (Grants No. 2308085QA34).
X.W. and C.B. are supported by Australian Research Council grants DP210101636, DP220100643 and LE21010001.
\end{acknowledgments}

\bibliography{ref.bib}

\begin{thebibliography}{70}%
\makeatletter
\providecommand \@ifxundefined [1]{%
 \@ifx{#1\undefined}
}%
\providecommand \@ifnum [1]{%
 \ifnum #1\expandafter \@firstoftwo
 \else \expandafter \@secondoftwo
 \fi
}%
\providecommand \@ifx [1]{%
 \ifx #1\expandafter \@firstoftwo
 \else \expandafter \@secondoftwo
 \fi
}%
\providecommand \natexlab [1]{#1}%
\providecommand \enquote  [1]{``#1''}%
\providecommand \bibnamefont  [1]{#1}%
\providecommand \bibfnamefont [1]{#1}%
\providecommand \citenamefont [1]{#1}%
\providecommand \href@noop [0]{\@secondoftwo}%
\providecommand \href [0]{\begingroup \@sanitize@url \@href}%
\providecommand \@href[1]{\@@startlink{#1}\@@href}%
\providecommand \@@href[1]{\endgroup#1\@@endlink}%
\providecommand \@sanitize@url [0]{\catcode `\\12\catcode `\$12\catcode `\&12\catcode `\#12\catcode `\^12\catcode `\_12\catcode `\%12\relax}%
\providecommand \@@startlink[1]{}%
\providecommand \@@endlink[0]{}%
\providecommand \url  [0]{\begingroup\@sanitize@url \@url }%
\providecommand \@url [1]{\endgroup\@href {#1}{\urlprefix }}%
\providecommand \urlprefix  [0]{URL }%
\providecommand \Eprint [0]{\href }%
\providecommand \doibase [0]{http://dx.doi.org/}%
\providecommand \selectlanguage [0]{\@gobble}%
\providecommand \bibinfo  [0]{\@secondoftwo}%
\providecommand \bibfield  [0]{\@secondoftwo}%
\providecommand \translation [1]{[#1]}%
\providecommand \BibitemOpen [0]{}%
\providecommand \bibitemStop [0]{}%
\providecommand \bibitemNoStop [0]{.\EOS\space}%
\providecommand \EOS [0]{\spacefactor3000\relax}%
\providecommand \BibitemShut  [1]{\csname bibitem#1\endcsname}%
\let\auto@bib@innerbib\@empty
\bibitem [{\citenamefont {Abbott}\ \emph {et~al.}(2016)\citenamefont {Abbott} \emph {et~al.}}]{LIGOScientific:2016aoc}%
  \BibitemOpen
  \bibfield  {author} {\bibinfo {author} {\bibfnamefont {B.~P.}\ \bibnamefont {Abbott}} \emph {et~al.} (\bibinfo {collaboration} {LIGO Scientific, Virgo}),\ }\bibfield  {title} {\enquote {\bibinfo {title} {{Observation of Gravitational Waves from a Binary Black Hole Merger}},}\ }\href {\doibase 10.1103/PhysRevLett.116.061102} {\bibfield  {journal} {\bibinfo  {journal} {Phys. Rev. Lett.}\ }\textbf {\bibinfo {volume} {116}},\ \bibinfo {pages} {061102} (\bibinfo {year} {2016})},\ \Eprint {http://arxiv.org/abs/1602.03837} {arXiv:1602.03837 [gr-qc]} \BibitemShut {NoStop}%
\bibitem [{\citenamefont {Agazie}\ \emph {et~al.}(2023)\citenamefont {Agazie} \emph {et~al.}}]{NANOGrav:2023gor}%
  \BibitemOpen
  \bibfield  {author} {\bibinfo {author} {\bibfnamefont {Gabriella}\ \bibnamefont {Agazie}} \emph {et~al.} (\bibinfo {collaboration} {NANOGrav}),\ }\bibfield  {title} {\enquote {\bibinfo {title} {{The NANOGrav 15 yr Data Set: Evidence for a Gravitational-wave Background}},}\ }\href {\doibase 10.3847/2041-8213/acdac6} {\bibfield  {journal} {\bibinfo  {journal} {Astrophys. J. Lett.}\ }\textbf {\bibinfo {volume} {951}},\ \bibinfo {pages} {L8} (\bibinfo {year} {2023})},\ \Eprint {http://arxiv.org/abs/2306.16213} {arXiv:2306.16213 [astro-ph.HE]} \BibitemShut {NoStop}%
\bibitem [{\citenamefont {Xu}\ \emph {et~al.}(2023)\citenamefont {Xu} \emph {et~al.}}]{Xu:2023wog}%
  \BibitemOpen
  \bibfield  {author} {\bibinfo {author} {\bibfnamefont {Heng}\ \bibnamefont {Xu}} \emph {et~al.},\ }\bibfield  {title} {\enquote {\bibinfo {title} {{Searching for the Nano-Hertz Stochastic Gravitational Wave Background with the Chinese Pulsar Timing Array Data Release I}},}\ }\href {\doibase 10.1088/1674-4527/acdfa5} {\bibfield  {journal} {\bibinfo  {journal} {Res. Astron. Astrophys.}\ }\textbf {\bibinfo {volume} {23}},\ \bibinfo {pages} {075024} (\bibinfo {year} {2023})},\ \Eprint {http://arxiv.org/abs/2306.16216} {arXiv:2306.16216 [astro-ph.HE]} \BibitemShut {NoStop}%
\bibitem [{\citenamefont {Antoniadis}\ \emph {et~al.}(2023)\citenamefont {Antoniadis} \emph {et~al.}}]{EPTA:2023fyk}%
  \BibitemOpen
  \bibfield  {author} {\bibinfo {author} {\bibfnamefont {J.}~\bibnamefont {Antoniadis}} \emph {et~al.} (\bibinfo {collaboration} {EPTA, InPTA:}),\ }\bibfield  {title} {\enquote {\bibinfo {title} {{The second data release from the European Pulsar Timing Array - III. Search for gravitational wave signals}},}\ }\href {\doibase 10.1051/0004-6361/202346844} {\bibfield  {journal} {\bibinfo  {journal} {Astron. Astrophys.}\ }\textbf {\bibinfo {volume} {678}},\ \bibinfo {pages} {A50} (\bibinfo {year} {2023})},\ \Eprint {http://arxiv.org/abs/2306.16214} {arXiv:2306.16214 [astro-ph.HE]} \BibitemShut {NoStop}%
\bibitem [{\citenamefont {Reardon}\ \emph {et~al.}(2023)\citenamefont {Reardon} \emph {et~al.}}]{Reardon:2023gzh}%
  \BibitemOpen
  \bibfield  {author} {\bibinfo {author} {\bibfnamefont {Daniel~J.}\ \bibnamefont {Reardon}} \emph {et~al.},\ }\bibfield  {title} {\enquote {\bibinfo {title} {{Search for an Isotropic Gravitational-wave Background with the Parkes Pulsar Timing Array}},}\ }\href {\doibase 10.3847/2041-8213/acdd02} {\bibfield  {journal} {\bibinfo  {journal} {Astrophys. J. Lett.}\ }\textbf {\bibinfo {volume} {951}},\ \bibinfo {pages} {L6} (\bibinfo {year} {2023})},\ \Eprint {http://arxiv.org/abs/2306.16215} {arXiv:2306.16215 [astro-ph.HE]} \BibitemShut {NoStop}%
\bibitem [{\citenamefont {Amaro-Seoane}\ \emph {et~al.}(2017)\citenamefont {Amaro-Seoane} \emph {et~al.}}]{LISA:2017pwj}%
  \BibitemOpen
  \bibfield  {author} {\bibinfo {author} {\bibfnamefont {Pau}\ \bibnamefont {Amaro-Seoane}} \emph {et~al.} (\bibinfo {collaboration} {LISA}),\ }\bibfield  {title} {\enquote {\bibinfo {title} {{Laser Interferometer Space Antenna}},}\ }\href@noop {} {\  (\bibinfo {year} {2017})},\ \Eprint {http://arxiv.org/abs/1702.00786} {arXiv:1702.00786 [astro-ph.IM]} \BibitemShut {NoStop}%
\bibitem [{\citenamefont {Colpi}\ \emph {et~al.}(2024)\citenamefont {Colpi} \emph {et~al.}}]{Colpi:2024xhw}%
  \BibitemOpen
  \bibfield  {author} {\bibinfo {author} {\bibfnamefont {Monica}\ \bibnamefont {Colpi}} \emph {et~al.},\ }\bibfield  {title} {\enquote {\bibinfo {title} {{LISA Definition Study Report}},}\ }\href@noop {} {\  (\bibinfo {year} {2024})},\ \Eprint {http://arxiv.org/abs/2402.07571} {arXiv:2402.07571 [astro-ph.CO]} \BibitemShut {NoStop}%
\bibitem [{\citenamefont {Luo}\ \emph {et~al.}(2016)\citenamefont {Luo} \emph {et~al.}}]{TianQin:2015yph}%
  \BibitemOpen
  \bibfield  {author} {\bibinfo {author} {\bibfnamefont {Jun}\ \bibnamefont {Luo}} \emph {et~al.} (\bibinfo {collaboration} {TianQin}),\ }\bibfield  {title} {\enquote {\bibinfo {title} {{TianQin: a space-borne gravitational wave detector}},}\ }\href {\doibase 10.1088/0264-9381/33/3/035010} {\bibfield  {journal} {\bibinfo  {journal} {Class. Quant. Grav.}\ }\textbf {\bibinfo {volume} {33}},\ \bibinfo {pages} {035010} (\bibinfo {year} {2016})},\ \Eprint {http://arxiv.org/abs/1512.02076} {arXiv:1512.02076 [astro-ph.IM]} \BibitemShut {NoStop}%
\bibitem [{\citenamefont {Hu}\ and\ \citenamefont {Wu}(2017)}]{Hu:2017mde}%
  \BibitemOpen
  \bibfield  {author} {\bibinfo {author} {\bibfnamefont {Wen-Rui}\ \bibnamefont {Hu}}\ and\ \bibinfo {author} {\bibfnamefont {Yue-Liang}\ \bibnamefont {Wu}},\ }\bibfield  {title} {\enquote {\bibinfo {title} {{The Taiji Program in Space for gravitational wave physics and the nature of gravity}},}\ }\href {\doibase 10.1093/nsr/nwx116} {\bibfield  {journal} {\bibinfo  {journal} {Natl. Sci. Rev.}\ }\textbf {\bibinfo {volume} {4}},\ \bibinfo {pages} {685--686} (\bibinfo {year} {2017})}\BibitemShut {NoStop}%
\bibitem [{\citenamefont {Corbin}\ and\ \citenamefont {Cornish}(2006)}]{Corbin:2005ny}%
  \BibitemOpen
  \bibfield  {author} {\bibinfo {author} {\bibfnamefont {Vincent}\ \bibnamefont {Corbin}}\ and\ \bibinfo {author} {\bibfnamefont {Neil~J.}\ \bibnamefont {Cornish}},\ }\bibfield  {title} {\enquote {\bibinfo {title} {{Detecting the cosmic gravitational wave background with the big bang observer}},}\ }\href {\doibase 10.1088/0264-9381/23/7/014} {\bibfield  {journal} {\bibinfo  {journal} {Class. Quant. Grav.}\ }\textbf {\bibinfo {volume} {23}},\ \bibinfo {pages} {2435--2446} (\bibinfo {year} {2006})},\ \Eprint {http://arxiv.org/abs/gr-qc/0512039} {arXiv:gr-qc/0512039} \BibitemShut {NoStop}%
\bibitem [{\citenamefont {Kawamura}\ \emph {et~al.}(2011)\citenamefont {Kawamura} \emph {et~al.}}]{Kawamura:2011zz}%
  \BibitemOpen
  \bibfield  {author} {\bibinfo {author} {\bibfnamefont {Seiji}\ \bibnamefont {Kawamura}} \emph {et~al.},\ }\bibfield  {title} {\enquote {\bibinfo {title} {{The Japanese space gravitational wave antenna: DECIGO}},}\ }\href {\doibase 10.1088/0264-9381/28/9/094011} {\bibfield  {journal} {\bibinfo  {journal} {Class. Quant. Grav.}\ }\textbf {\bibinfo {volume} {28}},\ \bibinfo {pages} {094011} (\bibinfo {year} {2011})}\BibitemShut {NoStop}%
\bibitem [{\citenamefont {Kudoh}\ \emph {et~al.}(2006)\citenamefont {Kudoh}, \citenamefont {Taruya}, \citenamefont {Hiramatsu},\ and\ \citenamefont {Himemoto}}]{Kudoh:2005as}%
  \BibitemOpen
  \bibfield  {author} {\bibinfo {author} {\bibfnamefont {Hideaki}\ \bibnamefont {Kudoh}}, \bibinfo {author} {\bibfnamefont {Atsushi}\ \bibnamefont {Taruya}}, \bibinfo {author} {\bibfnamefont {Takashi}\ \bibnamefont {Hiramatsu}}, \ and\ \bibinfo {author} {\bibfnamefont {Yoshiaki}\ \bibnamefont {Himemoto}},\ }\bibfield  {title} {\enquote {\bibinfo {title} {{Detecting a gravitational-wave background with next-generation space interferometers}},}\ }\href {\doibase 10.1103/PhysRevD.73.064006} {\bibfield  {journal} {\bibinfo  {journal} {Phys. Rev. D}\ }\textbf {\bibinfo {volume} {73}},\ \bibinfo {pages} {064006} (\bibinfo {year} {2006})},\ \Eprint {http://arxiv.org/abs/gr-qc/0511145} {arXiv:gr-qc/0511145} \BibitemShut {NoStop}%
\bibitem [{\citenamefont {Caprini}\ \emph {et~al.}(2016)\citenamefont {Caprini} \emph {et~al.}}]{Caprini:2015zlo}%
  \BibitemOpen
  \bibfield  {author} {\bibinfo {author} {\bibfnamefont {Chiara}\ \bibnamefont {Caprini}} \emph {et~al.},\ }\bibfield  {title} {\enquote {\bibinfo {title} {{Science with the space-based interferometer eLISA. II: Gravitational waves from cosmological phase transitions}},}\ }\href {\doibase 10.1088/1475-7516/2016/04/001} {\bibfield  {journal} {\bibinfo  {journal} {JCAP}\ }\textbf {\bibinfo {volume} {04}},\ \bibinfo {pages} {001} (\bibinfo {year} {2016})},\ \Eprint {http://arxiv.org/abs/1512.06239} {arXiv:1512.06239 [astro-ph.CO]} \BibitemShut {NoStop}%
\bibitem [{\citenamefont {Caprini}\ \emph {et~al.}(2020)\citenamefont {Caprini} \emph {et~al.}}]{Caprini:2019egz}%
  \BibitemOpen
  \bibfield  {author} {\bibinfo {author} {\bibfnamefont {Chiara}\ \bibnamefont {Caprini}} \emph {et~al.},\ }\bibfield  {title} {\enquote {\bibinfo {title} {{Detecting gravitational waves from cosmological phase transitions with LISA: an update}},}\ }\href {\doibase 10.1088/1475-7516/2020/03/024} {\bibfield  {journal} {\bibinfo  {journal} {JCAP}\ }\textbf {\bibinfo {volume} {03}},\ \bibinfo {pages} {024} (\bibinfo {year} {2020})},\ \Eprint {http://arxiv.org/abs/1910.13125} {arXiv:1910.13125 [astro-ph.CO]} \BibitemShut {NoStop}%
\bibitem [{\citenamefont {Athron}\ \emph {et~al.}(2024)\citenamefont {Athron}, \citenamefont {Bal\'azs}, \citenamefont {Fowlie}, \citenamefont {Morris},\ and\ \citenamefont {Wu}}]{Athron:2023xlk}%
  \BibitemOpen
  \bibfield  {author} {\bibinfo {author} {\bibfnamefont {Peter}\ \bibnamefont {Athron}}, \bibinfo {author} {\bibfnamefont {Csaba}\ \bibnamefont {Bal\'azs}}, \bibinfo {author} {\bibfnamefont {Andrew}\ \bibnamefont {Fowlie}}, \bibinfo {author} {\bibfnamefont {Lachlan}\ \bibnamefont {Morris}}, \ and\ \bibinfo {author} {\bibfnamefont {Lei}\ \bibnamefont {Wu}},\ }\bibfield  {title} {\enquote {\bibinfo {title} {{Cosmological phase transitions: From perturbative particle physics to gravitational waves}},}\ }\href {\doibase 10.1016/j.ppnp.2023.104094} {\bibfield  {journal} {\bibinfo  {journal} {Prog. Part. Nucl. Phys.}\ }\textbf {\bibinfo {volume} {135}},\ \bibinfo {pages} {104094} (\bibinfo {year} {2024})},\ \Eprint {http://arxiv.org/abs/2305.02357} {arXiv:2305.02357 [hep-ph]} \BibitemShut {NoStop}%
\bibitem [{\citenamefont {Hindmarsh}\ \emph {et~al.}(2014)\citenamefont {Hindmarsh}, \citenamefont {Huber}, \citenamefont {Rummukainen},\ and\ \citenamefont {Weir}}]{Hindmarsh:2013xza}%
  \BibitemOpen
  \bibfield  {author} {\bibinfo {author} {\bibfnamefont {Mark}\ \bibnamefont {Hindmarsh}}, \bibinfo {author} {\bibfnamefont {Stephan~J.}\ \bibnamefont {Huber}}, \bibinfo {author} {\bibfnamefont {Kari}\ \bibnamefont {Rummukainen}}, \ and\ \bibinfo {author} {\bibfnamefont {David~J.}\ \bibnamefont {Weir}},\ }\bibfield  {title} {\enquote {\bibinfo {title} {{Gravitational waves from the sound of a first order phase transition}},}\ }\href {\doibase 10.1103/PhysRevLett.112.041301} {\bibfield  {journal} {\bibinfo  {journal} {Phys. Rev. Lett.}\ }\textbf {\bibinfo {volume} {112}},\ \bibinfo {pages} {041301} (\bibinfo {year} {2014})},\ \Eprint {http://arxiv.org/abs/1304.2433} {arXiv:1304.2433 [hep-ph]} \BibitemShut {NoStop}%
\bibitem [{\citenamefont {Hindmarsh}\ \emph {et~al.}(2015)\citenamefont {Hindmarsh}, \citenamefont {Huber}, \citenamefont {Rummukainen},\ and\ \citenamefont {Weir}}]{Hindmarsh:2015qta}%
  \BibitemOpen
  \bibfield  {author} {\bibinfo {author} {\bibfnamefont {Mark}\ \bibnamefont {Hindmarsh}}, \bibinfo {author} {\bibfnamefont {Stephan~J.}\ \bibnamefont {Huber}}, \bibinfo {author} {\bibfnamefont {Kari}\ \bibnamefont {Rummukainen}}, \ and\ \bibinfo {author} {\bibfnamefont {David~J.}\ \bibnamefont {Weir}},\ }\bibfield  {title} {\enquote {\bibinfo {title} {{Numerical simulations of acoustically generated gravitational waves at a first order phase transition}},}\ }\href {\doibase 10.1103/PhysRevD.92.123009} {\bibfield  {journal} {\bibinfo  {journal} {Phys. Rev. D}\ }\textbf {\bibinfo {volume} {92}},\ \bibinfo {pages} {123009} (\bibinfo {year} {2015})},\ \Eprint {http://arxiv.org/abs/1504.03291} {arXiv:1504.03291 [astro-ph.CO]} \BibitemShut {NoStop}%
\bibitem [{\citenamefont {Hindmarsh}\ \emph {et~al.}(2017)\citenamefont {Hindmarsh}, \citenamefont {Huber}, \citenamefont {Rummukainen},\ and\ \citenamefont {Weir}}]{Hindmarsh:2017gnf}%
  \BibitemOpen
  \bibfield  {author} {\bibinfo {author} {\bibfnamefont {Mark}\ \bibnamefont {Hindmarsh}}, \bibinfo {author} {\bibfnamefont {Stephan~J.}\ \bibnamefont {Huber}}, \bibinfo {author} {\bibfnamefont {Kari}\ \bibnamefont {Rummukainen}}, \ and\ \bibinfo {author} {\bibfnamefont {David~J.}\ \bibnamefont {Weir}},\ }\bibfield  {title} {\enquote {\bibinfo {title} {{Shape of the acoustic gravitational wave power spectrum from a first order phase transition}},}\ }\href {\doibase 10.1103/PhysRevD.96.103520} {\bibfield  {journal} {\bibinfo  {journal} {Phys. Rev. D}\ }\textbf {\bibinfo {volume} {96}},\ \bibinfo {pages} {103520} (\bibinfo {year} {2017})},\ \bibinfo {note} {[Erratum: Phys.Rev.D 101, 089902 (2020)]},\ \Eprint {http://arxiv.org/abs/1704.05871} {arXiv:1704.05871 [astro-ph.CO]} \BibitemShut {NoStop}%
\bibitem [{\citenamefont {Jinno}\ \emph {et~al.}(2021{\natexlab{a}})\citenamefont {Jinno}, \citenamefont {Konstandin},\ and\ \citenamefont {Rubira}}]{Jinno:2020eqg}%
  \BibitemOpen
  \bibfield  {author} {\bibinfo {author} {\bibfnamefont {Ryusuke}\ \bibnamefont {Jinno}}, \bibinfo {author} {\bibfnamefont {Thomas}\ \bibnamefont {Konstandin}}, \ and\ \bibinfo {author} {\bibfnamefont {Henrique}\ \bibnamefont {Rubira}},\ }\bibfield  {title} {\enquote {\bibinfo {title} {{A hybrid simulation of gravitational wave production in first-order phase transitions}},}\ }\href {\doibase 10.1088/1475-7516/2021/04/014} {\bibfield  {journal} {\bibinfo  {journal} {JCAP}\ }\textbf {\bibinfo {volume} {04}},\ \bibinfo {pages} {014} (\bibinfo {year} {2021}{\natexlab{a}})},\ \Eprint {http://arxiv.org/abs/2010.00971} {arXiv:2010.00971 [astro-ph.CO]} \BibitemShut {NoStop}%
\bibitem [{\citenamefont {Jinno}\ \emph {et~al.}(2023)\citenamefont {Jinno}, \citenamefont {Konstandin}, \citenamefont {Rubira},\ and\ \citenamefont {Stomberg}}]{Jinno:2022mie}%
  \BibitemOpen
  \bibfield  {author} {\bibinfo {author} {\bibfnamefont {Ryusuke}\ \bibnamefont {Jinno}}, \bibinfo {author} {\bibfnamefont {Thomas}\ \bibnamefont {Konstandin}}, \bibinfo {author} {\bibfnamefont {Henrique}\ \bibnamefont {Rubira}}, \ and\ \bibinfo {author} {\bibfnamefont {Isak}\ \bibnamefont {Stomberg}},\ }\bibfield  {title} {\enquote {\bibinfo {title} {{Higgsless simulations of cosmological phase transitions and gravitational waves}},}\ }\href {\doibase 10.1088/1475-7516/2023/02/011} {\bibfield  {journal} {\bibinfo  {journal} {JCAP}\ }\textbf {\bibinfo {volume} {02}},\ \bibinfo {pages} {011} (\bibinfo {year} {2023})},\ \Eprint {http://arxiv.org/abs/2209.04369} {arXiv:2209.04369 [astro-ph.CO]} \BibitemShut {NoStop}%
\bibitem [{\citenamefont {Hindmarsh}(2018)}]{Hindmarsh:2016lnk}%
  \BibitemOpen
  \bibfield  {author} {\bibinfo {author} {\bibfnamefont {Mark}\ \bibnamefont {Hindmarsh}},\ }\bibfield  {title} {\enquote {\bibinfo {title} {{Sound shell model for acoustic gravitational wave production at a first-order phase transition in the early Universe}},}\ }\href {\doibase 10.1103/PhysRevLett.120.071301} {\bibfield  {journal} {\bibinfo  {journal} {Phys. Rev. Lett.}\ }\textbf {\bibinfo {volume} {120}},\ \bibinfo {pages} {071301} (\bibinfo {year} {2018})},\ \Eprint {http://arxiv.org/abs/1608.04735} {arXiv:1608.04735 [astro-ph.CO]} \BibitemShut {NoStop}%
\bibitem [{\citenamefont {Hindmarsh}\ and\ \citenamefont {Hijazi}(2019)}]{Hindmarsh:2019phv}%
  \BibitemOpen
  \bibfield  {author} {\bibinfo {author} {\bibfnamefont {Mark}\ \bibnamefont {Hindmarsh}}\ and\ \bibinfo {author} {\bibfnamefont {Mulham}\ \bibnamefont {Hijazi}},\ }\bibfield  {title} {\enquote {\bibinfo {title} {{Gravitational waves from first order cosmological phase transitions in the Sound Shell Model}},}\ }\href {\doibase 10.1088/1475-7516/2019/12/062} {\bibfield  {journal} {\bibinfo  {journal} {JCAP}\ }\textbf {\bibinfo {volume} {12}},\ \bibinfo {pages} {062} (\bibinfo {year} {2019})},\ \Eprint {http://arxiv.org/abs/1909.10040} {arXiv:1909.10040 [astro-ph.CO]} \BibitemShut {NoStop}%
\bibitem [{\citenamefont {Roper~Pol}\ \emph {et~al.}(2024)\citenamefont {Roper~Pol}, \citenamefont {Procacci},\ and\ \citenamefont {Caprini}}]{RoperPol:2023dzg}%
  \BibitemOpen
  \bibfield  {author} {\bibinfo {author} {\bibfnamefont {Alberto}\ \bibnamefont {Roper~Pol}}, \bibinfo {author} {\bibfnamefont {Simona}\ \bibnamefont {Procacci}}, \ and\ \bibinfo {author} {\bibfnamefont {Chiara}\ \bibnamefont {Caprini}},\ }\bibfield  {title} {\enquote {\bibinfo {title} {{Characterization of the gravitational wave spectrum from sound waves within the sound shell model}},}\ }\href {\doibase 10.1103/PhysRevD.109.063531} {\bibfield  {journal} {\bibinfo  {journal} {Phys. Rev. D}\ }\textbf {\bibinfo {volume} {109}},\ \bibinfo {pages} {063531} (\bibinfo {year} {2024})},\ \Eprint {http://arxiv.org/abs/2308.12943} {arXiv:2308.12943 [gr-qc]} \BibitemShut {NoStop}%
\bibitem [{\citenamefont {Sharma}\ \emph {et~al.}(2023)\citenamefont {Sharma}, \citenamefont {Dahl}, \citenamefont {Brandenburg},\ and\ \citenamefont {Hindmarsh}}]{Sharma:2023mao}%
  \BibitemOpen
  \bibfield  {author} {\bibinfo {author} {\bibfnamefont {Ramkishor}\ \bibnamefont {Sharma}}, \bibinfo {author} {\bibfnamefont {Jani}\ \bibnamefont {Dahl}}, \bibinfo {author} {\bibfnamefont {Axel}\ \bibnamefont {Brandenburg}}, \ and\ \bibinfo {author} {\bibfnamefont {Mark}\ \bibnamefont {Hindmarsh}},\ }\bibfield  {title} {\enquote {\bibinfo {title} {{Shallow relic gravitational wave spectrum with acoustic peak}},}\ }\href {\doibase 10.1088/1475-7516/2023/12/042} {\bibfield  {journal} {\bibinfo  {journal} {JCAP}\ }\textbf {\bibinfo {volume} {12}},\ \bibinfo {pages} {042} (\bibinfo {year} {2023})},\ \Eprint {http://arxiv.org/abs/2308.12916} {arXiv:2308.12916 [gr-qc]} \BibitemShut {NoStop}%
\bibitem [{\citenamefont {Giombi}\ \emph {et~al.}(2024)\citenamefont {Giombi}, \citenamefont {Dahl},\ and\ \citenamefont {Hindmarsh}}]{Giombi:2024kju}%
  \BibitemOpen
  \bibfield  {author} {\bibinfo {author} {\bibfnamefont {Lorenzo}\ \bibnamefont {Giombi}}, \bibinfo {author} {\bibfnamefont {Jani}\ \bibnamefont {Dahl}}, \ and\ \bibinfo {author} {\bibfnamefont {Mark}\ \bibnamefont {Hindmarsh}},\ }\bibfield  {title} {\enquote {\bibinfo {title} {{Signatures of the speed of sound on the gravitational wave power spectrum from sound waves}},}\ }\href@noop {} {\  (\bibinfo {year} {2024})},\ \Eprint {http://arxiv.org/abs/2409.01426} {arXiv:2409.01426 [gr-qc]} \BibitemShut {NoStop}%
\bibitem [{\citenamefont {Jinno}\ \emph {et~al.}(2021{\natexlab{b}})\citenamefont {Jinno}, \citenamefont {Konstandin}, \citenamefont {Rubira},\ and\ \citenamefont {van~de Vis}}]{Jinno:2021ury}%
  \BibitemOpen
  \bibfield  {author} {\bibinfo {author} {\bibfnamefont {Ryusuke}\ \bibnamefont {Jinno}}, \bibinfo {author} {\bibfnamefont {Thomas}\ \bibnamefont {Konstandin}}, \bibinfo {author} {\bibfnamefont {Henrique}\ \bibnamefont {Rubira}}, \ and\ \bibinfo {author} {\bibfnamefont {Jorinde}\ \bibnamefont {van~de Vis}},\ }\bibfield  {title} {\enquote {\bibinfo {title} {{Effect of density fluctuations on gravitational wave production in first-order phase transitions}},}\ }\href {\doibase 10.1088/1475-7516/2021/12/019} {\bibfield  {journal} {\bibinfo  {journal} {JCAP}\ }\textbf {\bibinfo {volume} {12}},\ \bibinfo {pages} {019} (\bibinfo {year} {2021}{\natexlab{b}})},\ \Eprint {http://arxiv.org/abs/2108.11947} {arXiv:2108.11947 [astro-ph.CO]} \BibitemShut {NoStop}%
\bibitem [{\citenamefont {Guo}\ \emph {et~al.}(2021)\citenamefont {Guo}, \citenamefont {Sinha}, \citenamefont {Vagie},\ and\ \citenamefont {White}}]{Guo:2020grp}%
  \BibitemOpen
  \bibfield  {author} {\bibinfo {author} {\bibfnamefont {Huai-Ke}\ \bibnamefont {Guo}}, \bibinfo {author} {\bibfnamefont {Kuver}\ \bibnamefont {Sinha}}, \bibinfo {author} {\bibfnamefont {Daniel}\ \bibnamefont {Vagie}}, \ and\ \bibinfo {author} {\bibfnamefont {Graham}\ \bibnamefont {White}},\ }\bibfield  {title} {\enquote {\bibinfo {title} {{Phase Transitions in an Expanding Universe: Stochastic Gravitational Waves in Standard and Non-Standard Histories}},}\ }\href {\doibase 10.1088/1475-7516/2021/01/001} {\bibfield  {journal} {\bibinfo  {journal} {JCAP}\ }\textbf {\bibinfo {volume} {01}},\ \bibinfo {pages} {001} (\bibinfo {year} {2021})},\ \Eprint {http://arxiv.org/abs/2007.08537} {arXiv:2007.08537 [hep-ph]} \BibitemShut {NoStop}%
\bibitem [{\citenamefont {Wang}\ \emph {et~al.}(2022)\citenamefont {Wang}, \citenamefont {Huang},\ and\ \citenamefont {Li}}]{Wang:2021dwl}%
  \BibitemOpen
  \bibfield  {author} {\bibinfo {author} {\bibfnamefont {Xiao}\ \bibnamefont {Wang}}, \bibinfo {author} {\bibfnamefont {Fa~Peng}\ \bibnamefont {Huang}}, \ and\ \bibinfo {author} {\bibfnamefont {Yongping}\ \bibnamefont {Li}},\ }\bibfield  {title} {\enquote {\bibinfo {title} {{Sound velocity effects on the phase transition gravitational wave spectrum in the sound shell model}},}\ }\href {\doibase 10.1103/PhysRevD.105.103513} {\bibfield  {journal} {\bibinfo  {journal} {Phys. Rev. D}\ }\textbf {\bibinfo {volume} {105}},\ \bibinfo {pages} {103513} (\bibinfo {year} {2022})},\ \Eprint {http://arxiv.org/abs/2112.14650} {arXiv:2112.14650 [astro-ph.CO]} \BibitemShut {NoStop}%
\bibitem [{\citenamefont {Cai}\ \emph {et~al.}(2023)\citenamefont {Cai}, \citenamefont {Wang},\ and\ \citenamefont {Yuwen}}]{Cai:2023guc}%
  \BibitemOpen
  \bibfield  {author} {\bibinfo {author} {\bibfnamefont {Rong-Gen}\ \bibnamefont {Cai}}, \bibinfo {author} {\bibfnamefont {Shao-Jiang}\ \bibnamefont {Wang}}, \ and\ \bibinfo {author} {\bibfnamefont {Zi-Yan}\ \bibnamefont {Yuwen}},\ }\bibfield  {title} {\enquote {\bibinfo {title} {{Hydrodynamic sound shell model}},}\ }\href {\doibase 10.1103/PhysRevD.108.L021502} {\bibfield  {journal} {\bibinfo  {journal} {Phys. Rev. D}\ }\textbf {\bibinfo {volume} {108}},\ \bibinfo {pages} {L021502} (\bibinfo {year} {2023})},\ \Eprint {http://arxiv.org/abs/2305.00074} {arXiv:2305.00074 [gr-qc]} \BibitemShut {NoStop}%
\bibitem [{\citenamefont {Zhang}(1993)}]{Zhang:1992fs}%
  \BibitemOpen
  \bibfield  {author} {\bibinfo {author} {\bibfnamefont {Xin-min}\ \bibnamefont {Zhang}},\ }\bibfield  {title} {\enquote {\bibinfo {title} {{Operators analysis for Higgs potential and cosmological bound on Higgs mass}},}\ }\href {\doibase 10.1103/PhysRevD.47.3065} {\bibfield  {journal} {\bibinfo  {journal} {Phys. Rev. D}\ }\textbf {\bibinfo {volume} {47}},\ \bibinfo {pages} {3065--3067} (\bibinfo {year} {1993})},\ \Eprint {http://arxiv.org/abs/hep-ph/9301277} {arXiv:hep-ph/9301277} \BibitemShut {NoStop}%
\bibitem [{\citenamefont {Grojean}\ \emph {et~al.}(2005)\citenamefont {Grojean}, \citenamefont {Servant},\ and\ \citenamefont {Wells}}]{Grojean:2004xa}%
  \BibitemOpen
  \bibfield  {author} {\bibinfo {author} {\bibfnamefont {Christophe}\ \bibnamefont {Grojean}}, \bibinfo {author} {\bibfnamefont {Geraldine}\ \bibnamefont {Servant}}, \ and\ \bibinfo {author} {\bibfnamefont {James~D.}\ \bibnamefont {Wells}},\ }\bibfield  {title} {\enquote {\bibinfo {title} {{First-order electroweak phase transition in the standard model with a low cutoff}},}\ }\href {\doibase 10.1103/PhysRevD.71.036001} {\bibfield  {journal} {\bibinfo  {journal} {Phys. Rev. D}\ }\textbf {\bibinfo {volume} {71}},\ \bibinfo {pages} {036001} (\bibinfo {year} {2005})},\ \Eprint {http://arxiv.org/abs/hep-ph/0407019} {arXiv:hep-ph/0407019} \BibitemShut {NoStop}%
\bibitem [{\citenamefont {Chung}\ \emph {et~al.}(2013)\citenamefont {Chung}, \citenamefont {Long},\ and\ \citenamefont {Wang}}]{Chung:2012vg}%
  \BibitemOpen
  \bibfield  {author} {\bibinfo {author} {\bibfnamefont {Daniel J.~H.}\ \bibnamefont {Chung}}, \bibinfo {author} {\bibfnamefont {Andrew~J.}\ \bibnamefont {Long}}, \ and\ \bibinfo {author} {\bibfnamefont {Lian-Tao}\ \bibnamefont {Wang}},\ }\bibfield  {title} {\enquote {\bibinfo {title} {{125 GeV Higgs boson and electroweak phase transition model classes}},}\ }\href {\doibase 10.1103/PhysRevD.87.023509} {\bibfield  {journal} {\bibinfo  {journal} {Phys. Rev. D}\ }\textbf {\bibinfo {volume} {87}},\ \bibinfo {pages} {023509} (\bibinfo {year} {2013})},\ \Eprint {http://arxiv.org/abs/1209.1819} {arXiv:1209.1819 [hep-ph]} \BibitemShut {NoStop}%
\bibitem [{\citenamefont {Huang}\ \emph {et~al.}(2016)\citenamefont {Huang}, \citenamefont {Wan}, \citenamefont {Wang}, \citenamefont {Cai},\ and\ \citenamefont {Zhang}}]{Huang:2016odd}%
  \BibitemOpen
  \bibfield  {author} {\bibinfo {author} {\bibfnamefont {Fa~Peng}\ \bibnamefont {Huang}}, \bibinfo {author} {\bibfnamefont {Youping}\ \bibnamefont {Wan}}, \bibinfo {author} {\bibfnamefont {Dong-Gang}\ \bibnamefont {Wang}}, \bibinfo {author} {\bibfnamefont {Yi-Fu}\ \bibnamefont {Cai}}, \ and\ \bibinfo {author} {\bibfnamefont {Xinmin}\ \bibnamefont {Zhang}},\ }\bibfield  {title} {\enquote {\bibinfo {title} {{Hearing the echoes of electroweak baryogenesis with gravitational wave detectors}},}\ }\href {\doibase 10.1103/PhysRevD.94.041702} {\bibfield  {journal} {\bibinfo  {journal} {Phys. Rev. D}\ }\textbf {\bibinfo {volume} {94}},\ \bibinfo {pages} {041702} (\bibinfo {year} {2016})},\ \Eprint {http://arxiv.org/abs/1601.01640} {arXiv:1601.01640 [hep-ph]} \BibitemShut {NoStop}%
\bibitem [{\citenamefont {Wang}\ \emph {et~al.}(2020{\natexlab{a}})\citenamefont {Wang}, \citenamefont {Huang},\ and\ \citenamefont {Zhang}}]{Wang:2020jrd}%
  \BibitemOpen
  \bibfield  {author} {\bibinfo {author} {\bibfnamefont {Xiao}\ \bibnamefont {Wang}}, \bibinfo {author} {\bibfnamefont {Fa~Peng}\ \bibnamefont {Huang}}, \ and\ \bibinfo {author} {\bibfnamefont {Xinmin}\ \bibnamefont {Zhang}},\ }\bibfield  {title} {\enquote {\bibinfo {title} {{Phase transition dynamics and gravitational wave spectra of strong first-order phase transition in supercooled universe}},}\ }\href {\doibase 10.1088/1475-7516/2020/05/045} {\bibfield  {journal} {\bibinfo  {journal} {JCAP}\ }\textbf {\bibinfo {volume} {05}},\ \bibinfo {pages} {045} (\bibinfo {year} {2020}{\natexlab{a}})},\ \Eprint {http://arxiv.org/abs/2003.08892} {arXiv:2003.08892 [hep-ph]} \BibitemShut {NoStop}%
\bibitem [{\citenamefont {Espinosa}\ and\ \citenamefont {Quiros}(2007)}]{Espinosa:2007qk}%
  \BibitemOpen
  \bibfield  {author} {\bibinfo {author} {\bibfnamefont {Jose~Ramon}\ \bibnamefont {Espinosa}}\ and\ \bibinfo {author} {\bibfnamefont {Mariano}\ \bibnamefont {Quiros}},\ }\bibfield  {title} {\enquote {\bibinfo {title} {{Novel Effects in Electroweak Breaking from a Hidden Sector}},}\ }\href {\doibase 10.1103/PhysRevD.76.076004} {\bibfield  {journal} {\bibinfo  {journal} {Phys. Rev. D}\ }\textbf {\bibinfo {volume} {76}},\ \bibinfo {pages} {076004} (\bibinfo {year} {2007})},\ \Eprint {http://arxiv.org/abs/hep-ph/0701145} {arXiv:hep-ph/0701145} \BibitemShut {NoStop}%
\bibitem [{\citenamefont {Profumo}\ \emph {et~al.}(2007)\citenamefont {Profumo}, \citenamefont {Ramsey-Musolf},\ and\ \citenamefont {Shaughnessy}}]{Profumo:2007wc}%
  \BibitemOpen
  \bibfield  {author} {\bibinfo {author} {\bibfnamefont {Stefano}\ \bibnamefont {Profumo}}, \bibinfo {author} {\bibfnamefont {Michael~J.}\ \bibnamefont {Ramsey-Musolf}}, \ and\ \bibinfo {author} {\bibfnamefont {Gabe}\ \bibnamefont {Shaughnessy}},\ }\bibfield  {title} {\enquote {\bibinfo {title} {{Singlet Higgs phenomenology and the electroweak phase transition}},}\ }\href {\doibase 10.1088/1126-6708/2007/08/010} {\bibfield  {journal} {\bibinfo  {journal} {JHEP}\ }\textbf {\bibinfo {volume} {08}},\ \bibinfo {pages} {010} (\bibinfo {year} {2007})},\ \Eprint {http://arxiv.org/abs/0705.2425} {arXiv:0705.2425 [hep-ph]} \BibitemShut {NoStop}%
\bibitem [{\citenamefont {Espinosa}\ \emph {et~al.}(2012)\citenamefont {Espinosa}, \citenamefont {Konstandin},\ and\ \citenamefont {Riva}}]{Espinosa:2011ax}%
  \BibitemOpen
  \bibfield  {author} {\bibinfo {author} {\bibfnamefont {Jose~R.}\ \bibnamefont {Espinosa}}, \bibinfo {author} {\bibfnamefont {Thomas}\ \bibnamefont {Konstandin}}, \ and\ \bibinfo {author} {\bibfnamefont {Francesco}\ \bibnamefont {Riva}},\ }\bibfield  {title} {\enquote {\bibinfo {title} {{Strong Electroweak Phase Transitions in the Standard Model with a Singlet}},}\ }\href {\doibase 10.1016/j.nuclphysb.2011.09.010} {\bibfield  {journal} {\bibinfo  {journal} {Nucl. Phys. B}\ }\textbf {\bibinfo {volume} {854}},\ \bibinfo {pages} {592--630} (\bibinfo {year} {2012})},\ \Eprint {http://arxiv.org/abs/1107.5441} {arXiv:1107.5441 [hep-ph]} \BibitemShut {NoStop}%
\bibitem [{\citenamefont {Chiang}\ \emph {et~al.}(2019)\citenamefont {Chiang}, \citenamefont {Li},\ and\ \citenamefont {Senaha}}]{Chiang:2018gsn}%
  \BibitemOpen
  \bibfield  {author} {\bibinfo {author} {\bibfnamefont {Cheng-Wei}\ \bibnamefont {Chiang}}, \bibinfo {author} {\bibfnamefont {Yen-Ting}\ \bibnamefont {Li}}, \ and\ \bibinfo {author} {\bibfnamefont {Eibun}\ \bibnamefont {Senaha}},\ }\bibfield  {title} {\enquote {\bibinfo {title} {{Revisiting electroweak phase transition in the standard model with a real singlet scalar}},}\ }\href {\doibase 10.1016/j.physletb.2018.12.017} {\bibfield  {journal} {\bibinfo  {journal} {Phys. Lett. B}\ }\textbf {\bibinfo {volume} {789}},\ \bibinfo {pages} {154--159} (\bibinfo {year} {2019})},\ \Eprint {http://arxiv.org/abs/1808.01098} {arXiv:1808.01098 [hep-ph]} \BibitemShut {NoStop}%
\bibitem [{\citenamefont {Athron}\ \emph {et~al.}(2023)\citenamefont {Athron}, \citenamefont {Balazs}, \citenamefont {Fowlie}, \citenamefont {Morris}, \citenamefont {White},\ and\ \citenamefont {Zhang}}]{Athron:2022jyi}%
  \BibitemOpen
  \bibfield  {author} {\bibinfo {author} {\bibfnamefont {Peter}\ \bibnamefont {Athron}}, \bibinfo {author} {\bibfnamefont {Csaba}\ \bibnamefont {Balazs}}, \bibinfo {author} {\bibfnamefont {Andrew}\ \bibnamefont {Fowlie}}, \bibinfo {author} {\bibfnamefont {Lachlan}\ \bibnamefont {Morris}}, \bibinfo {author} {\bibfnamefont {Graham}\ \bibnamefont {White}}, \ and\ \bibinfo {author} {\bibfnamefont {Yang}\ \bibnamefont {Zhang}},\ }\bibfield  {title} {\enquote {\bibinfo {title} {{How arbitrary are perturbative calculations of the electroweak phase transition?}}}\ }\href {\doibase 10.1007/JHEP01(2023)050} {\bibfield  {journal} {\bibinfo  {journal} {JHEP}\ }\textbf {\bibinfo {volume} {01}},\ \bibinfo {pages} {050} (\bibinfo {year} {2023})},\ \Eprint {http://arxiv.org/abs/2208.01319} {arXiv:2208.01319 [hep-ph]} \BibitemShut {NoStop}%
\bibitem [{\citenamefont {Lee}(1973)}]{Lee:1973iz}%
  \BibitemOpen
  \bibfield  {author} {\bibinfo {author} {\bibfnamefont {T.~D.}\ \bibnamefont {Lee}},\ }\bibfield  {title} {\enquote {\bibinfo {title} {{A Theory of Spontaneous T Violation}},}\ }\href {\doibase 10.1103/PhysRevD.8.1226} {\bibfield  {journal} {\bibinfo  {journal} {Phys. Rev. D}\ }\textbf {\bibinfo {volume} {8}},\ \bibinfo {pages} {1226--1239} (\bibinfo {year} {1973})}\BibitemShut {NoStop}%
\bibitem [{\citenamefont {Branco}\ \emph {et~al.}(2012)\citenamefont {Branco}, \citenamefont {Ferreira}, \citenamefont {Lavoura}, \citenamefont {Rebelo}, \citenamefont {Sher},\ and\ \citenamefont {Silva}}]{Branco:2011iw}%
  \BibitemOpen
  \bibfield  {author} {\bibinfo {author} {\bibfnamefont {G.~C.}\ \bibnamefont {Branco}}, \bibinfo {author} {\bibfnamefont {P.~M.}\ \bibnamefont {Ferreira}}, \bibinfo {author} {\bibfnamefont {L.}~\bibnamefont {Lavoura}}, \bibinfo {author} {\bibfnamefont {M.~N.}\ \bibnamefont {Rebelo}}, \bibinfo {author} {\bibfnamefont {Marc}\ \bibnamefont {Sher}}, \ and\ \bibinfo {author} {\bibfnamefont {Joao~P.}\ \bibnamefont {Silva}},\ }\bibfield  {title} {\enquote {\bibinfo {title} {{Theory and phenomenology of two-Higgs-doublet models}},}\ }\href {\doibase 10.1016/j.physrep.2012.02.002} {\bibfield  {journal} {\bibinfo  {journal} {Phys. Rept.}\ }\textbf {\bibinfo {volume} {516}},\ \bibinfo {pages} {1--102} (\bibinfo {year} {2012})},\ \Eprint {http://arxiv.org/abs/1106.0034} {arXiv:1106.0034 [hep-ph]} \BibitemShut {NoStop}%
\bibitem [{\citenamefont {Dorsch}\ \emph {et~al.}(2017)\citenamefont {Dorsch}, \citenamefont {Huber}, \citenamefont {Konstandin},\ and\ \citenamefont {No}}]{Dorsch:2016nrg}%
  \BibitemOpen
  \bibfield  {author} {\bibinfo {author} {\bibfnamefont {G.~C.}\ \bibnamefont {Dorsch}}, \bibinfo {author} {\bibfnamefont {S.~J.}\ \bibnamefont {Huber}}, \bibinfo {author} {\bibfnamefont {T.}~\bibnamefont {Konstandin}}, \ and\ \bibinfo {author} {\bibfnamefont {J.~M.}\ \bibnamefont {No}},\ }\bibfield  {title} {\enquote {\bibinfo {title} {{A Second Higgs Doublet in the Early Universe: Baryogenesis and Gravitational Waves}},}\ }\href {\doibase 10.1088/1475-7516/2017/05/052} {\bibfield  {journal} {\bibinfo  {journal} {JCAP}\ }\textbf {\bibinfo {volume} {05}},\ \bibinfo {pages} {052} (\bibinfo {year} {2017})},\ \Eprint {http://arxiv.org/abs/1611.05874} {arXiv:1611.05874 [hep-ph]} \BibitemShut {NoStop}%
\bibitem [{\citenamefont {Basler}\ \emph {et~al.}(2018)\citenamefont {Basler}, \citenamefont {M\"uhlleitner},\ and\ \citenamefont {Wittbrodt}}]{Basler:2017uxn}%
  \BibitemOpen
  \bibfield  {author} {\bibinfo {author} {\bibfnamefont {Philipp}\ \bibnamefont {Basler}}, \bibinfo {author} {\bibfnamefont {Margarete}\ \bibnamefont {M\"uhlleitner}}, \ and\ \bibinfo {author} {\bibfnamefont {Jonas}\ \bibnamefont {Wittbrodt}},\ }\bibfield  {title} {\enquote {\bibinfo {title} {{The CP-Violating 2HDM in Light of a Strong First Order Electroweak Phase Transition and Implications for Higgs Pair Production}},}\ }\href {\doibase 10.1007/JHEP03(2018)061} {\bibfield  {journal} {\bibinfo  {journal} {JHEP}\ }\textbf {\bibinfo {volume} {03}},\ \bibinfo {pages} {061} (\bibinfo {year} {2018})},\ \Eprint {http://arxiv.org/abs/1711.04097} {arXiv:1711.04097 [hep-ph]} \BibitemShut {NoStop}%
\bibitem [{\citenamefont {Fontes}\ \emph {et~al.}(2018)\citenamefont {Fontes}, \citenamefont {M\"uhlleitner}, \citenamefont {Rom\~ao}, \citenamefont {Santos}, \citenamefont {Silva},\ and\ \citenamefont {Wittbrodt}}]{Fontes:2017zfn}%
  \BibitemOpen
  \bibfield  {author} {\bibinfo {author} {\bibfnamefont {Duarte}\ \bibnamefont {Fontes}}, \bibinfo {author} {\bibfnamefont {Margarete}\ \bibnamefont {M\"uhlleitner}}, \bibinfo {author} {\bibfnamefont {Jorge~C.}\ \bibnamefont {Rom\~ao}}, \bibinfo {author} {\bibfnamefont {Rui}\ \bibnamefont {Santos}}, \bibinfo {author} {\bibfnamefont {Jo\~ao~P.}\ \bibnamefont {Silva}}, \ and\ \bibinfo {author} {\bibfnamefont {Jonas}\ \bibnamefont {Wittbrodt}},\ }\bibfield  {title} {\enquote {\bibinfo {title} {{The C2HDM revisited}},}\ }\href {\doibase 10.1007/JHEP02(2018)073} {\bibfield  {journal} {\bibinfo  {journal} {JHEP}\ }\textbf {\bibinfo {volume} {02}},\ \bibinfo {pages} {073} (\bibinfo {year} {2018})},\ \Eprint {http://arxiv.org/abs/1711.09419} {arXiv:1711.09419 [hep-ph]} \BibitemShut {NoStop}%
\bibitem [{\citenamefont {Wang}\ \emph {et~al.}(2020{\natexlab{b}})\citenamefont {Wang}, \citenamefont {Huang},\ and\ \citenamefont {Zhang}}]{Wang:2019pet}%
  \BibitemOpen
  \bibfield  {author} {\bibinfo {author} {\bibfnamefont {Xiao}\ \bibnamefont {Wang}}, \bibinfo {author} {\bibfnamefont {Fa~Peng}\ \bibnamefont {Huang}}, \ and\ \bibinfo {author} {\bibfnamefont {Xinmin}\ \bibnamefont {Zhang}},\ }\bibfield  {title} {\enquote {\bibinfo {title} {{Gravitational wave and collider signals in complex two-Higgs doublet model with dynamical CP-violation at finite temperature}},}\ }\href {\doibase 10.1103/PhysRevD.101.015015} {\bibfield  {journal} {\bibinfo  {journal} {Phys. Rev. D}\ }\textbf {\bibinfo {volume} {101}},\ \bibinfo {pages} {015015} (\bibinfo {year} {2020}{\natexlab{b}})},\ \Eprint {http://arxiv.org/abs/1909.02978} {arXiv:1909.02978 [hep-ph]} \BibitemShut {NoStop}%
\bibitem [{\citenamefont {Laine}\ and\ \citenamefont {Vuorinen}(2016)}]{Laine:2016hma}%
  \BibitemOpen
  \bibfield  {author} {\bibinfo {author} {\bibfnamefont {Mikko}\ \bibnamefont {Laine}}\ and\ \bibinfo {author} {\bibfnamefont {Aleksi}\ \bibnamefont {Vuorinen}},\ }\href {\doibase 10.1007/978-3-319-31933-9} {\emph {\bibinfo {title} {{Basics of Thermal Field Theory}}}},\ Vol.\ \bibinfo {volume} {925}\ (\bibinfo  {publisher} {Springer},\ \bibinfo {year} {2016})\ \Eprint {http://arxiv.org/abs/1701.01554} {arXiv:1701.01554 [hep-ph]} \BibitemShut {NoStop}%
\bibitem [{\citenamefont {Enqvist}\ \emph {et~al.}(1992)\citenamefont {Enqvist}, \citenamefont {Ignatius}, \citenamefont {Kajantie},\ and\ \citenamefont {Rummukainen}}]{Enqvist:1991xw}%
  \BibitemOpen
  \bibfield  {author} {\bibinfo {author} {\bibfnamefont {K.}~\bibnamefont {Enqvist}}, \bibinfo {author} {\bibfnamefont {J.}~\bibnamefont {Ignatius}}, \bibinfo {author} {\bibfnamefont {K.}~\bibnamefont {Kajantie}}, \ and\ \bibinfo {author} {\bibfnamefont {K.}~\bibnamefont {Rummukainen}},\ }\bibfield  {title} {\enquote {\bibinfo {title} {{Nucleation and bubble growth in a first order cosmological electroweak phase transition}},}\ }\href {\doibase 10.1103/PhysRevD.45.3415} {\bibfield  {journal} {\bibinfo  {journal} {Phys. Rev. D}\ }\textbf {\bibinfo {volume} {45}},\ \bibinfo {pages} {3415--3428} (\bibinfo {year} {1992})}\BibitemShut {NoStop}%
\bibitem [{\citenamefont {Moore}\ and\ \citenamefont {Prokopec}(1995{\natexlab{a}})}]{Moore:1995ua}%
  \BibitemOpen
  \bibfield  {author} {\bibinfo {author} {\bibfnamefont {Guy~D.}\ \bibnamefont {Moore}}\ and\ \bibinfo {author} {\bibfnamefont {Tomislav}\ \bibnamefont {Prokopec}},\ }\bibfield  {title} {\enquote {\bibinfo {title} {{Bubble wall velocity in a first order electroweak phase transition}},}\ }\href {\doibase 10.1103/PhysRevLett.75.777} {\bibfield  {journal} {\bibinfo  {journal} {Phys. Rev. Lett.}\ }\textbf {\bibinfo {volume} {75}},\ \bibinfo {pages} {777--780} (\bibinfo {year} {1995}{\natexlab{a}})},\ \Eprint {http://arxiv.org/abs/hep-ph/9503296} {arXiv:hep-ph/9503296} \BibitemShut {NoStop}%
\bibitem [{\citenamefont {Moore}\ and\ \citenamefont {Prokopec}(1995{\natexlab{b}})}]{Moore:1995si}%
  \BibitemOpen
  \bibfield  {author} {\bibinfo {author} {\bibfnamefont {Guy~D.}\ \bibnamefont {Moore}}\ and\ \bibinfo {author} {\bibfnamefont {Tomislav}\ \bibnamefont {Prokopec}},\ }\bibfield  {title} {\enquote {\bibinfo {title} {{How fast can the wall move? A Study of the electroweak phase transition dynamics}},}\ }\href {\doibase 10.1103/PhysRevD.52.7182} {\bibfield  {journal} {\bibinfo  {journal} {Phys. Rev. D}\ }\textbf {\bibinfo {volume} {52}},\ \bibinfo {pages} {7182--7204} (\bibinfo {year} {1995}{\natexlab{b}})},\ \Eprint {http://arxiv.org/abs/hep-ph/9506475} {arXiv:hep-ph/9506475} \BibitemShut {NoStop}%
\bibitem [{\citenamefont {Laurent}\ and\ \citenamefont {Cline}(2020)}]{Laurent:2020gpg}%
  \BibitemOpen
  \bibfield  {author} {\bibinfo {author} {\bibfnamefont {Benoit}\ \bibnamefont {Laurent}}\ and\ \bibinfo {author} {\bibfnamefont {James~M.}\ \bibnamefont {Cline}},\ }\bibfield  {title} {\enquote {\bibinfo {title} {{Fluid equations for fast-moving electroweak bubble walls}},}\ }\href {\doibase 10.1103/PhysRevD.102.063516} {\bibfield  {journal} {\bibinfo  {journal} {Phys. Rev. D}\ }\textbf {\bibinfo {volume} {102}},\ \bibinfo {pages} {063516} (\bibinfo {year} {2020})},\ \Eprint {http://arxiv.org/abs/2007.10935} {arXiv:2007.10935 [hep-ph]} \BibitemShut {NoStop}%
\bibitem [{\citenamefont {Laurent}\ and\ \citenamefont {Cline}(2022)}]{Laurent:2022jrs}%
  \BibitemOpen
  \bibfield  {author} {\bibinfo {author} {\bibfnamefont {Benoit}\ \bibnamefont {Laurent}}\ and\ \bibinfo {author} {\bibfnamefont {James~M.}\ \bibnamefont {Cline}},\ }\bibfield  {title} {\enquote {\bibinfo {title} {{First principles determination of bubble wall velocity}},}\ }\href {\doibase 10.1103/PhysRevD.106.023501} {\bibfield  {journal} {\bibinfo  {journal} {Phys. Rev. D}\ }\textbf {\bibinfo {volume} {106}},\ \bibinfo {pages} {023501} (\bibinfo {year} {2022})},\ \Eprint {http://arxiv.org/abs/2204.13120} {arXiv:2204.13120 [hep-ph]} \BibitemShut {NoStop}%
\bibitem [{\citenamefont {Wang}\ \emph {et~al.}(2020{\natexlab{c}})\citenamefont {Wang}, \citenamefont {Huang},\ and\ \citenamefont {Zhang}}]{Wang:2020zlf}%
  \BibitemOpen
  \bibfield  {author} {\bibinfo {author} {\bibfnamefont {Xiao}\ \bibnamefont {Wang}}, \bibinfo {author} {\bibfnamefont {Fa~Peng}\ \bibnamefont {Huang}}, \ and\ \bibinfo {author} {\bibfnamefont {Xinmin}\ \bibnamefont {Zhang}},\ }\bibfield  {title} {\enquote {\bibinfo {title} {{Bubble wall velocity beyond leading-log approximation in electroweak phase transition}},}\ }\href@noop {} {\  (\bibinfo {year} {2020}{\natexlab{c}})},\ \Eprint {http://arxiv.org/abs/2011.12903} {arXiv:2011.12903 [hep-ph]} \BibitemShut {NoStop}%
\bibitem [{\citenamefont {Jiang}\ \emph {et~al.}(2023)\citenamefont {Jiang}, \citenamefont {Huang},\ and\ \citenamefont {Wang}}]{Jiang:2022btc}%
  \BibitemOpen
  \bibfield  {author} {\bibinfo {author} {\bibfnamefont {Siyu}\ \bibnamefont {Jiang}}, \bibinfo {author} {\bibfnamefont {Fa~Peng}\ \bibnamefont {Huang}}, \ and\ \bibinfo {author} {\bibfnamefont {Xiao}\ \bibnamefont {Wang}},\ }\bibfield  {title} {\enquote {\bibinfo {title} {{Bubble wall velocity during electroweak phase transition in the inert doublet model}},}\ }\href {\doibase 10.1103/PhysRevD.107.095005} {\bibfield  {journal} {\bibinfo  {journal} {Phys. Rev. D}\ }\textbf {\bibinfo {volume} {107}},\ \bibinfo {pages} {095005} (\bibinfo {year} {2023})},\ \Eprint {http://arxiv.org/abs/2211.13142} {arXiv:2211.13142 [hep-ph]} \BibitemShut {NoStop}%
\bibitem [{\citenamefont {Boyd}(2013)}]{boyd2013chebyshev}%
  \BibitemOpen
  \bibfield  {author} {\bibinfo {author} {\bibfnamefont {J.P.}\ \bibnamefont {Boyd}},\ }\href {https://books.google.com.au/books?id=b4TCAgAAQBAJ} {\emph {\bibinfo {title} {Chebyshev and Fourier Spectral Methods: Second Revised Edition}}},\ Dover Books on Mathematics\ (\bibinfo  {publisher} {Dover Publications},\ \bibinfo {year} {2013})\BibitemShut {NoStop}%
\bibitem [{\citenamefont {Espinosa}\ \emph {et~al.}(2010)\citenamefont {Espinosa}, \citenamefont {Konstandin}, \citenamefont {No},\ and\ \citenamefont {Servant}}]{Espinosa:2010hh}%
  \BibitemOpen
  \bibfield  {author} {\bibinfo {author} {\bibfnamefont {Jose~R.}\ \bibnamefont {Espinosa}}, \bibinfo {author} {\bibfnamefont {Thomas}\ \bibnamefont {Konstandin}}, \bibinfo {author} {\bibfnamefont {Jose~M.}\ \bibnamefont {No}}, \ and\ \bibinfo {author} {\bibfnamefont {Geraldine}\ \bibnamefont {Servant}},\ }\bibfield  {title} {\enquote {\bibinfo {title} {{Energy Budget of Cosmological First-order Phase Transitions}},}\ }\href {\doibase 10.1088/1475-7516/2010/06/028} {\bibfield  {journal} {\bibinfo  {journal} {JCAP}\ }\textbf {\bibinfo {volume} {06}},\ \bibinfo {pages} {028} (\bibinfo {year} {2010})},\ \Eprint {http://arxiv.org/abs/1004.4187} {arXiv:1004.4187 [hep-ph]} \BibitemShut {NoStop}%
\bibitem [{\citenamefont {Wang}\ \emph {et~al.}(2021)\citenamefont {Wang}, \citenamefont {Huang},\ and\ \citenamefont {Zhang}}]{Wang:2020nzm}%
  \BibitemOpen
  \bibfield  {author} {\bibinfo {author} {\bibfnamefont {Xiao}\ \bibnamefont {Wang}}, \bibinfo {author} {\bibfnamefont {Fa~Peng}\ \bibnamefont {Huang}}, \ and\ \bibinfo {author} {\bibfnamefont {Xinmin}\ \bibnamefont {Zhang}},\ }\bibfield  {title} {\enquote {\bibinfo {title} {{Energy budget and the gravitational wave spectra beyond the bag model}},}\ }\href {\doibase 10.1103/PhysRevD.103.103520} {\bibfield  {journal} {\bibinfo  {journal} {Phys. Rev. D}\ }\textbf {\bibinfo {volume} {103}},\ \bibinfo {pages} {103520} (\bibinfo {year} {2021})},\ \Eprint {http://arxiv.org/abs/2010.13770} {arXiv:2010.13770 [astro-ph.CO]} \BibitemShut {NoStop}%
\bibitem [{\citenamefont {Tian}\ \emph {et~al.}()\citenamefont {Tian}, \citenamefont {Wang},\ and\ \citenamefont {Bal\'azs}}]{Long:2024}%
  \BibitemOpen
  \bibfield  {author} {\bibinfo {author} {\bibfnamefont {Chi}\ \bibnamefont {Tian}}, \bibinfo {author} {\bibfnamefont {Xiao}\ \bibnamefont {Wang}}, \ and\ \bibinfo {author} {\bibfnamefont {Csaba}\ \bibnamefont {Bal\'azs}},\ }\bibfield  {title} {\enquote {\bibinfo {title} {{Gravitational waves from cosmological first-order phase transitions with precise hydrodynamics}},}\ }\href@noop {} {\ }\Eprint {http://arxiv.org/abs/In preparation} {In preparation} \BibitemShut {NoStop}%
\bibitem [{\citenamefont {Kurganov}\ and\ \citenamefont {Tadmor}(2000)}]{Kurganov:2000ovy}%
  \BibitemOpen
  \bibfield  {author} {\bibinfo {author} {\bibfnamefont {Alexander}\ \bibnamefont {Kurganov}}\ and\ \bibinfo {author} {\bibfnamefont {Eitan}\ \bibnamefont {Tadmor}},\ }\bibfield  {title} {\enquote {\bibinfo {title} {{New High-Resolution Central Schemes for Nonlinear Conservation Laws and Convection\textendash{}Diffusion Equations}},}\ }\href {\doibase 10.1006/jcph.2000.6459} {\bibfield  {journal} {\bibinfo  {journal} {J. Comput. Phys.}\ }\textbf {\bibinfo {volume} {160}},\ \bibinfo {pages} {241--282} (\bibinfo {year} {2000})}\BibitemShut {NoStop}%
\bibitem [{\citenamefont {Wang}\ \emph {et~al.}(2023)\citenamefont {Wang}, \citenamefont {Tian},\ and\ \citenamefont {Huang}}]{Wang:2023jto}%
  \BibitemOpen
  \bibfield  {author} {\bibinfo {author} {\bibfnamefont {Xiao}\ \bibnamefont {Wang}}, \bibinfo {author} {\bibfnamefont {Chi}\ \bibnamefont {Tian}}, \ and\ \bibinfo {author} {\bibfnamefont {Fa~Peng}\ \bibnamefont {Huang}},\ }\bibfield  {title} {\enquote {\bibinfo {title} {{Model-dependent analysis method for energy budget of the cosmological first-order phase transition}},}\ }\href {\doibase 10.1088/1475-7516/2023/07/006} {\bibfield  {journal} {\bibinfo  {journal} {JCAP}\ }\textbf {\bibinfo {volume} {07}},\ \bibinfo {pages} {006} (\bibinfo {year} {2023})},\ \Eprint {http://arxiv.org/abs/2301.12328} {arXiv:2301.12328 [hep-ph]} \BibitemShut {NoStop}%
\bibitem [{\citenamefont {Thrane}\ and\ \citenamefont {Romano}(2013)}]{Thrane:2013oya}%
  \BibitemOpen
  \bibfield  {author} {\bibinfo {author} {\bibfnamefont {Eric}\ \bibnamefont {Thrane}}\ and\ \bibinfo {author} {\bibfnamefont {Joseph~D.}\ \bibnamefont {Romano}},\ }\bibfield  {title} {\enquote {\bibinfo {title} {{Sensitivity curves for searches for gravitational-wave backgrounds}},}\ }\href {\doibase 10.1103/PhysRevD.88.124032} {\bibfield  {journal} {\bibinfo  {journal} {Phys. Rev. D}\ }\textbf {\bibinfo {volume} {88}},\ \bibinfo {pages} {124032} (\bibinfo {year} {2013})},\ \Eprint {http://arxiv.org/abs/1310.5300} {arXiv:1310.5300 [astro-ph.IM]} \BibitemShut {NoStop}%
\bibitem [{\citenamefont {Schmitz}(2021)}]{Schmitz:2020syl}%
  \BibitemOpen
  \bibfield  {author} {\bibinfo {author} {\bibfnamefont {Kai}\ \bibnamefont {Schmitz}},\ }\bibfield  {title} {\enquote {\bibinfo {title} {{New Sensitivity Curves for Gravitational-Wave Signals from Cosmological Phase Transitions}},}\ }\href {\doibase 10.1007/JHEP01(2021)097} {\bibfield  {journal} {\bibinfo  {journal} {JHEP}\ }\textbf {\bibinfo {volume} {01}},\ \bibinfo {pages} {097} (\bibinfo {year} {2021})},\ \Eprint {http://arxiv.org/abs/2002.04615} {arXiv:2002.04615 [hep-ph]} \BibitemShut {NoStop}%
\bibitem [{\citenamefont {Ellis}\ \emph {et~al.}(2019)\citenamefont {Ellis}, \citenamefont {Lewicki}, \citenamefont {No},\ and\ \citenamefont {Vaskonen}}]{Ellis:2019oqb}%
  \BibitemOpen
  \bibfield  {author} {\bibinfo {author} {\bibfnamefont {John}\ \bibnamefont {Ellis}}, \bibinfo {author} {\bibfnamefont {Marek}\ \bibnamefont {Lewicki}}, \bibinfo {author} {\bibfnamefont {Jos\'e~Miguel}\ \bibnamefont {No}}, \ and\ \bibinfo {author} {\bibfnamefont {Ville}\ \bibnamefont {Vaskonen}},\ }\bibfield  {title} {\enquote {\bibinfo {title} {{Gravitational wave energy budget in strongly supercooled phase transitions}},}\ }\href {\doibase 10.1088/1475-7516/2019/06/024} {\bibfield  {journal} {\bibinfo  {journal} {JCAP}\ }\textbf {\bibinfo {volume} {06}},\ \bibinfo {pages} {024} (\bibinfo {year} {2019})},\ \Eprint {http://arxiv.org/abs/1903.09642} {arXiv:1903.09642 [hep-ph]} \BibitemShut {NoStop}%
\bibitem [{\citenamefont {Ellis}\ \emph {et~al.}(2020)\citenamefont {Ellis}, \citenamefont {Lewicki},\ and\ \citenamefont {Vaskonen}}]{Ellis:2020nnr}%
  \BibitemOpen
  \bibfield  {author} {\bibinfo {author} {\bibfnamefont {John}\ \bibnamefont {Ellis}}, \bibinfo {author} {\bibfnamefont {Marek}\ \bibnamefont {Lewicki}}, \ and\ \bibinfo {author} {\bibfnamefont {Ville}\ \bibnamefont {Vaskonen}},\ }\bibfield  {title} {\enquote {\bibinfo {title} {{Updated predictions for gravitational waves produced in a strongly supercooled phase transition}},}\ }\href {\doibase 10.1088/1475-7516/2020/11/020} {\bibfield  {journal} {\bibinfo  {journal} {JCAP}\ }\textbf {\bibinfo {volume} {11}},\ \bibinfo {pages} {020} (\bibinfo {year} {2020})},\ \Eprint {http://arxiv.org/abs/2007.15586} {arXiv:2007.15586 [astro-ph.CO]} \BibitemShut {NoStop}%
\bibitem [{\citenamefont {Linde}(1983)}]{Linde:1981zj}%
  \BibitemOpen
  \bibfield  {author} {\bibinfo {author} {\bibfnamefont {Andrei~D.}\ \bibnamefont {Linde}},\ }\bibfield  {title} {\enquote {\bibinfo {title} {{Decay of the False Vacuum at Finite Temperature}},}\ }\href {\doibase 10.1016/0550-3213(83)90072-X} {\bibfield  {journal} {\bibinfo  {journal} {Nucl. Phys. B}\ }\textbf {\bibinfo {volume} {216}},\ \bibinfo {pages} {421} (\bibinfo {year} {1983})},\ \bibinfo {note} {[Erratum: Nucl.Phys.B 223, 544 (1983)]}\BibitemShut {NoStop}%
\bibitem [{\citenamefont {Croon}\ \emph {et~al.}(2021)\citenamefont {Croon}, \citenamefont {Gould}, \citenamefont {Schicho}, \citenamefont {Tenkanen},\ and\ \citenamefont {White}}]{Croon:2020cgk}%
  \BibitemOpen
  \bibfield  {author} {\bibinfo {author} {\bibfnamefont {Djuna}\ \bibnamefont {Croon}}, \bibinfo {author} {\bibfnamefont {Oliver}\ \bibnamefont {Gould}}, \bibinfo {author} {\bibfnamefont {Philipp}\ \bibnamefont {Schicho}}, \bibinfo {author} {\bibfnamefont {Tuomas V.~I.}\ \bibnamefont {Tenkanen}}, \ and\ \bibinfo {author} {\bibfnamefont {Graham}\ \bibnamefont {White}},\ }\bibfield  {title} {\enquote {\bibinfo {title} {{Theoretical uncertainties for cosmological first-order phase transitions}},}\ }\href {\doibase 10.1007/JHEP04(2021)055} {\bibfield  {journal} {\bibinfo  {journal} {JHEP}\ }\textbf {\bibinfo {volume} {04}},\ \bibinfo {pages} {055} (\bibinfo {year} {2021})},\ \Eprint {http://arxiv.org/abs/2009.10080} {arXiv:2009.10080 [hep-ph]} \BibitemShut {NoStop}%
\bibitem [{\citenamefont {Schicho}\ \emph {et~al.}(2022)\citenamefont {Schicho}, \citenamefont {Tenkanen},\ and\ \citenamefont {White}}]{Schicho:2022wty}%
  \BibitemOpen
  \bibfield  {author} {\bibinfo {author} {\bibfnamefont {Philipp}\ \bibnamefont {Schicho}}, \bibinfo {author} {\bibfnamefont {Tuomas V.~I.}\ \bibnamefont {Tenkanen}}, \ and\ \bibinfo {author} {\bibfnamefont {Graham}\ \bibnamefont {White}},\ }\bibfield  {title} {\enquote {\bibinfo {title} {{Combining thermal resummation and gauge invariance for electroweak phase transition}},}\ }\href {\doibase 10.1007/JHEP11(2022)047} {\bibfield  {journal} {\bibinfo  {journal} {JHEP}\ }\textbf {\bibinfo {volume} {11}},\ \bibinfo {pages} {047} (\bibinfo {year} {2022})},\ \Eprint {http://arxiv.org/abs/2203.04284} {arXiv:2203.04284 [hep-ph]} \BibitemShut {NoStop}%
\bibitem [{\citenamefont {Gould}\ and\ \citenamefont {Tenkanen}(2024)}]{Gould:2023ovu}%
  \BibitemOpen
  \bibfield  {author} {\bibinfo {author} {\bibfnamefont {Oliver}\ \bibnamefont {Gould}}\ and\ \bibinfo {author} {\bibfnamefont {Tuomas V.~I.}\ \bibnamefont {Tenkanen}},\ }\bibfield  {title} {\enquote {\bibinfo {title} {{Perturbative effective field theory expansions for cosmological phase transitions}},}\ }\href {\doibase 10.1007/JHEP01(2024)048} {\bibfield  {journal} {\bibinfo  {journal} {JHEP}\ }\textbf {\bibinfo {volume} {01}},\ \bibinfo {pages} {048} (\bibinfo {year} {2024})},\ \Eprint {http://arxiv.org/abs/2309.01672} {arXiv:2309.01672 [hep-ph]} \BibitemShut {NoStop}%
\bibitem [{\citenamefont {Ekstedt}\ \emph {et~al.}(2024)\citenamefont {Ekstedt}, \citenamefont {Schicho},\ and\ \citenamefont {Tenkanen}}]{Ekstedt:2024etx}%
  \BibitemOpen
  \bibfield  {author} {\bibinfo {author} {\bibfnamefont {Andreas}\ \bibnamefont {Ekstedt}}, \bibinfo {author} {\bibfnamefont {Philipp}\ \bibnamefont {Schicho}}, \ and\ \bibinfo {author} {\bibfnamefont {Tuomas V.~I.}\ \bibnamefont {Tenkanen}},\ }\bibfield  {title} {\enquote {\bibinfo {title} {{Cosmological phase transitions at three loops: the final verdict on perturbation theory}},}\ }\href@noop {} {\  (\bibinfo {year} {2024})},\ \Eprint {http://arxiv.org/abs/2405.18349} {arXiv:2405.18349 [hep-ph]} \BibitemShut {NoStop}%
\bibitem [{\citenamefont {Farakos}\ \emph {et~al.}(1994)\citenamefont {Farakos}, \citenamefont {Kajantie}, \citenamefont {Rummukainen},\ and\ \citenamefont {Shaposhnikov}}]{Farakos:1994kx}%
  \BibitemOpen
  \bibfield  {author} {\bibinfo {author} {\bibfnamefont {K.}~\bibnamefont {Farakos}}, \bibinfo {author} {\bibfnamefont {K.}~\bibnamefont {Kajantie}}, \bibinfo {author} {\bibfnamefont {K.}~\bibnamefont {Rummukainen}}, \ and\ \bibinfo {author} {\bibfnamefont {Mikhail~E.}\ \bibnamefont {Shaposhnikov}},\ }\bibfield  {title} {\enquote {\bibinfo {title} {{3-D physics and the electroweak phase transition: Perturbation theory}},}\ }\href {\doibase 10.1016/0550-3213(94)90173-2} {\bibfield  {journal} {\bibinfo  {journal} {Nucl. Phys. B}\ }\textbf {\bibinfo {volume} {425}},\ \bibinfo {pages} {67--109} (\bibinfo {year} {1994})},\ \Eprint {http://arxiv.org/abs/hep-ph/9404201} {arXiv:hep-ph/9404201} \BibitemShut {NoStop}%
\bibitem [{\citenamefont {Kajantie}\ \emph {et~al.}(1996)\citenamefont {Kajantie}, \citenamefont {Laine}, \citenamefont {Rummukainen},\ and\ \citenamefont {Shaposhnikov}}]{Kajantie:1995dw}%
  \BibitemOpen
  \bibfield  {author} {\bibinfo {author} {\bibfnamefont {K.}~\bibnamefont {Kajantie}}, \bibinfo {author} {\bibfnamefont {M.}~\bibnamefont {Laine}}, \bibinfo {author} {\bibfnamefont {K.}~\bibnamefont {Rummukainen}}, \ and\ \bibinfo {author} {\bibfnamefont {Mikhail~E.}\ \bibnamefont {Shaposhnikov}},\ }\bibfield  {title} {\enquote {\bibinfo {title} {{Generic rules for high temperature dimensional reduction and their application to the standard model}},}\ }\href {\doibase 10.1016/0550-3213(95)00549-8} {\bibfield  {journal} {\bibinfo  {journal} {Nucl. Phys. B}\ }\textbf {\bibinfo {volume} {458}},\ \bibinfo {pages} {90--136} (\bibinfo {year} {1996})},\ \Eprint {http://arxiv.org/abs/hep-ph/9508379} {arXiv:hep-ph/9508379} \BibitemShut {NoStop}%
\end{thebibliography}%

\end{document}